\documentclass[a4paper,notoc]{JHEP3}
\usepackage{amsmath,amssymb,calc,latexsym}
\usepackage{graphicx,epsfig}

\newcommand{\tr}{\mathrm{Tr}}
\newcommand{\deriv}{\mathrm{d}}

\title{Radiative Corrections with 5D Mixed\\
Position-/Momentum-space Propagators}
\author{Martin Puchwein and Zoltan Kunszt\\
Institute for Theoretical Physics, ETH, CH-8093 Z\"urich, Switzerland\\
E-Mail: \email{puchwein@itp.phys.ethz.ch}, \email{kunszt@itp.phys.ethz.ch}}

\abstract{In higher dimensional field theories with compactified dimensions 
there are three  standard ways to do perturbative calculations: i) by
the summation over Kaluza-Klein towers; ii) by the summation over
 winding numbers
making use  of the Poisson-resummation formula and iii) 
by using mixed propagators, where the coordinates of
the four infinite dimensions are Fourier-transformed to momentum space 
while those of the compactified dimensions
are kept in configuration space.
 The third method is 
 broadly  used in 
 finite temperature field theory calculations. One of its advantage is
 that one can easily  separate 
  the  ultraviolet divergent terms  of 
 the uncompactified theory from the 
 non-local finite  corrections arising from windings around the
 compact dimensions. In this note   
 we demonstrate the use of this formalism by 
  calculating    one-loop self-energy corrections 
in a 5D theory formulated on the manifold
   ${\cal M}_4\,\otimes\, S_1$ and on the
orbifold ${\cal M}_4\,\otimes\ S_1/Z_2$.}
\begin{document}
\section{Introduction}
\label{sec:intro}
In the recent years field theories with additional compactified space-time dimensions have been proposed to shed new light on
the long standing problems of particle physics such as the unification with gravity and the hierarchy problem. Motivated
by string theory and the discovery of D-branes the possibility of gauge theories living in higher dimensions arose \cite{Horava:1996qa}. Unlike
string theory with its 10 dimensions, these theories live in the world-volume of a D-brane and can have as little as only
one compactified dimension. It has been shown that there exists the possibility that the scale of compactification can be
as low as $\mathrm{TeV}^{-1}$ \cite{Antoniadis:1990ew}, therefore being in the reach of the next generation particle
accelerators \cite{Arkani-Hamed:1998nn}.\\
In 4D field theory these models are described by an infinite number of Kaluza-Klein modes, corresponding to
states with discrete momenta in the compactified dimensions.
 To every field corresponds an infinite "tower"
of particles with the same quantum numbers and mass gaps the size of the compactification scale, therefore 
these higher dimensional field theories are in general unrenormalizeable.\\
Compactification on a circle (or respectively on a torus when one has more than one additional dimension) is most straightforward, but brings certain difficulties. It was soon realized that depending on the number of dimensions it
may not be possible to construct a theory that is chiral from the four-dimensional point of view \cite{Witten:1983ux}.
In the case of only
one additional dimension compactified on a circle it is indeed not possible to find such a theory. One way to overcome this apparent incompatibility is by the introduction of a discrete group acting non-freely on the 
compact manifold \cite{Dixon:1985jw,Dixon:1986jc}. The resulting manifold (orbifold) has singularities at the fixed points of the group action. These
points describe hypersurfaces, which are commonly referred to as "branes", but should not be confused with
D-branes. Imposing this new symmetry on the fields by assigning quantum numbers projects out those modes that have the "wrong" parities. For fermion fields one finds that the left- and right-handed components of the Dirac spinor have opposite charges under the symmetry. For the lowest lying mode this means that only one chirality remains in the spectrum. This leads to a low energy effective theory that is chiral.\\
In several papers \cite{Cheng:2002iz,vonGersdorff:2002as} 1-loop radiative corrections for a number of 5D models have
been calculated. Although the theories in question are non-renormalizeable, it has been found that when summing over
the contributions from all the KK-modes some physical observable quantities remain manifestly finite at 1-loop order. Since a non-renormalizeable
theory only makes sense as an effective theory one may ask whether summing over all KK-modes up
to arbitrary high energies is a legitimate thing to do \cite{Ghilencea:2001ug}, but this subtlety is well understood
by now \cite{Delgado:2001ex,Contino:2001gz,Barbieri:2001dm}. In this paper we want to present yet
another point of view to illuminate this topic, by working not in the picture of KK-modes but by treating the compactified dimension in
position space (for earlier work see \cite{Arkani-Hamed:1999za}). In the mixed formalism
the source of the infinities and their possible cancellation is more direct and evident than when working in
Fourier-space.\\
In Sec.(\ref{sec:S1}) we discuss the formalism of mixed momentum-/configuration-space for a theory compactified
on $S_2$ and calculate the
propagators for scalars, fermions and gauge bosons. We discuss the origin and possible cancellation of divergencies.
The proceeding section contains as an application the calculation of 1-loop corrections to the masses of the photon's
KK-tower in 5D QED. We
reproduce the results of \cite{Cheng:2002iz} and discuss why the corrections are finite. The more general case of
a non-abelian gauge theory in 5D can be found in the appendix. Going over to a more realistic theory we discuss
orbifold compactification in Sec.(\ref{sec:orb}) and redo the calculation of Sec.(\ref{sec:1ls1}) in this setup. The
conclusion discusses the results, new insights and outlooks of using the mixed momentum-/configuration-space approach.
\section{Propagators on $\mathbb{R}^4 \times S_1$}
\label{sec:S1}

We  consider  consider 
gauge  theories coupled to matter fields 
with one extra dimension compactified on a circle with
radius $R$. Such theories are always vector-like and therefore cannot serve as a realistic model, but
have to be considered as a toy model. 
Due to the compactification $5D$-Lorentz invariance is broken down to
the $4D$-Lorentz invariance of the uncompactified dimensions. Since $5D$ theories are in general non-renormalizeable,
they only make sense in the
context of effective field theories. Nevertheless one can make meaningful calculations as long as the
cutoff dependence is weak. This is the case if the typical energies are much lower than the cutoff and the quantity
in question is UV-insensitive. Also we have to specify a
regularization scheme to define the theory. When gauge invariance is of importance it is advantageous to work in dimensional regularization \cite{GrootNibbelink:2001bx}, which in odd space-time dimensions has the special feature that 1-loop diagrams are finite and $1/\epsilon$-poles only occur at
2-loop level \cite{Candelas:1984ae}. This is due to the fact that dimensional regularization captures only logarithmic
divergencies which are absent at 1-loop level (there are however linear divergencies, that do not show up). To estimate
the degree of divergence it will sometimes also be useful to make a sharp momentum cutoff.
To begin with we calculate the propagators of elementary fields in the mixed momentum/position space described in the
introduction.

\subsection{Scalar propagator}
\label{sec:S1_scal_prop} 
To start with we take a free theory of a scalar field defined on the manifold $\mathbb{R}^4 \times S_1$, with metric $g_{A B}=
\mathrm{diag}[1,-1,-1,-1,-1]$ and parameterize the extra dimension by $y\in [0,2\pi R[$. From the Lagrangian
\begin{equation}
\mathcal{L}=\frac{1}{2} \partial_A \phi\, \partial^A \phi - \frac{1}{2} m^2 \phi^2=\frac{1}{2} \partial_\mu \phi\, \partial^\mu \phi - \frac{1}{2}
 (\partial_5 \phi)^2 - \frac{1}{2} m^2 \phi^2\ ,
\end{equation}
one gets the equation of motion
\begin{displaymath}
\big(-\Box  + (\partial_5)^2 - m^2\big) \phi =0\ .
\end{displaymath}
The propagator is a Green's function of the above operator satisfying (in $d$ dimensions)
\begin{equation}
\big(-\Box  + (\partial_5)^2 - m^2\big) S_F(x-x',y,y')=i \delta^{(d-1)}(x-x') \delta(y-y')\label{eq:kg1}\ .
\end{equation}
In the uncompactified theory on $\mathbb{R}^5$ we already know the solution to this equation
\begin{displaymath}
S_F(x-x',y,y')=\int\frac{\deriv^{d} p}{(2\pi)^d} \frac{i}{p^2 -(p^5)^2-m^2 +i \epsilon} e^{-i(p\cdot (x-x') - p^5 (y-y'))}\ .
\end{displaymath}
Here $p^5$ is the momentum in the single compactified dimension and $p$ denotes the momentum in $d-1$ uncompactified
dimensions.
With this input it is easy to write down a propagator that fulfills the additional requirement of periodicity\footnote{This
is very similar to the procedure familiar from finite-temperature field theory \cite{Collins:1984xc} where $\beta=1/T$ plays
the role of the compactification radius. The only difference is that here we compactify a spacelike dimension, in contrast
to the timelike dimension compactified in thermal field theory.}.
\begin{equation}
S_F^c (x-x',y,y')= \sum_{n=-\infty}^{+\infty} S_F(x-x',y + 2 \pi R n,y')\label{eq:comprop}
\end{equation}
The sum on the right is a sum over winding modes, with $S_F(x-x',y + 2 \pi R n,y')$ winding $n$ times around the circle.
It can be evaluated by carrying out the $p^5$ integral (introducing $\chi=\sqrt{p^2-m^2+i\epsilon}$)
\begin{multline}
S_F^c (x-x',y,y')\\=\sum_{n=-\infty}^{+\infty} \int\frac{\deriv^{d-1} p}{(2\pi)^{d-1}} \int\frac{\deriv p^5}{2\pi}\frac{i}{p^2 -(p^5)^2-m^2 +i \epsilon} e^{-i(p\cdot (x-x') - p^5 (y+2 \pi R n -y'))}\\=
\sum_{n=-\infty}^{+\infty} \int\frac{\deriv^{d-1} p}{(2\pi)^{d-1}} \frac{e^{i \chi |y-y'+
2 \pi R n|}}{2 \chi} e^{-i p\cdot (x-x')}\ .\label{eq:nsum}
\end{multline}
On the compactified manifold $y,y' \in [0,2\pi R[$ the summation gives
\begin{equation}
S_F^c (x-x',y,y')=\int\frac{\deriv^{d-1} p}{(2\pi)^{d-1}} \frac{i \cos \chi (\pi R-|y-y'|)}{2 \chi \sin \chi \pi R}
e^{-i p\cdot (x-x')}\ .\label{eq:fullprop}
\end{equation}
However we will find that it is advantageous to write the propagator in yet another form, namely writing the
$n=0$ contribution to Eq.(\ref{eq:nsum}) explicitly. 
\begin{equation*}
S_F^c (x-x',y,y')= S_F(x-x',y,y')+S_F^{\mathrm{analy}}(x-x',y,y')
\end{equation*}
$S_F$ is just the propagator of the uncompactified theory and $S_F^{\mathrm{analy.}}$ is an analytical function that
is finite even for $y\rightarrow y'$ given by
\begin{multline}
S_F^{\mathrm{analy}}(x-x',y,y')=\sum_{\substack{n=-\infty\\n\neq 0}}^{+\infty} S_F(x-x',y + 2 \pi R n,y')\\
=\int\frac{\deriv^{d-1} p}{(2\pi)^{d-1}} \frac{e^{i \chi |y-y'|}+e^{-i \chi |y-y'|}}{2 \chi (e^{-i \chi (2\pi R) }-1)}
e^{-i p\cdot (x-x')}\ .\label{eq:propan}
\end{multline}
After a Wick-rotation with $\chi = i \chi_E$ (due to the $i \epsilon$ term) we find that for large Euclidean momenta
\begin{equation}
\frac{ e^{-\chi_E |y-y'|} + e^{\chi_E |y-y'|}}{2 \chi_E( e^{\chi_E (2 \pi R)}-1)}\ \rightarrow\ 
\frac{1}{2 \chi_E\, e^{\chi_E (2\pi R-|y-y'|)}} \qquad \mathrm{for}\quad \chi_E \gg R^{-1}\ .
\end{equation}
Using that $|y-y'|<2 \pi R$ we conclude that:\vspace{3mm}\\
$S_F^{\mathrm{analy}}$ is exponentially damped at high Euclidean momentum therefore rendering loop-integrals
that contain $S_F^{\mathrm{analy}}$ finite. Because of this, UV-divergencies occur only when one picks up a term $S_F(x-x',y,y')$ from every propagator in a loop integral.
\vspace{3mm}\\ In the limit of the radius going to infinity (going back to an infinite dimension) $S_F^{\mathrm{analy}}\rightarrow 0$.
This is what one would expect, since those terms in the propagator that correspond to a winding around the circle
cannot contribute in this limit. The winding number zero term goes over to the corresponding expression of the
uncompactified theory and is exactly the one that gives UV-divergencies. The counterterms of the uncompactified
and the compactified theory are the same, which is clear because the UV-divergencies are short-distance effects and do not feel the
compactification being a long distance effect. Again this is the same situation as encountered in finite temperature field theory, where the UV-divergencies
of the $T\neq0$ theory are exactly those of the $T=0$ theory.
\vspace{3mm}.

\subsection{Fermion propagator}
Next we calculate the propagator of a massive fermion field with Lagrangian
\begin{equation}
\mathcal{L}=i\, \bar\psi \gamma^A \partial_A \psi - m\, \bar\psi \psi=i\, \bar\psi \gamma^\mu \partial_\mu \psi +i\,\bar\psi \gamma^5 \partial_5 \psi-m\,\bar\psi \psi\ .
\end{equation}
Choosing the representation of the Dirac algebra to be
\begin{equation}
\gamma^\mu=\left(%
\begin{array}{cc}
  0 & \sigma^\mu \\
  \bar{\sigma}^\mu & 0 \\
\end{array}\right)\ ,\quad \gamma^5=\left(%
\begin{array}{cc}
  -i & 0 \\
 0 & i \\
\end{array}%
\right)\quad\mathrm{fulfilling:}\quad \{\gamma^A,\gamma^B\}=2 g^{A B}\ ,
\end{equation}
the Dirac equation is given by
\begin{equation}
\left(\begin{array}{cc}
\partial_5 -m & i\, \sigma^\mu \partial_\mu\\
i\, \bar\sigma^\mu \partial_\mu & -\partial_5 -m\\
\end{array}
\right)\psi =0\quad \Rightarrow \quad \big(\Box-(\partial_5)^2+m^2\big)\psi =0\label{eq:s1_direq}\ .
\end{equation}
It is straightforward to check explicitly that for $y,y'\in[0,2\pi R[$
\begin{equation}
S_F^c(x-x',y,y')=\int\!\! \frac{\deriv^4 p}{(2\pi)^4} e^{-ip.(x-x')}
\big(p_\mu \gamma^\mu +i \gamma^5\partial_5 + m\big) \frac{i \cos \chi(\pi R-|y-y'|)}{2 \chi \sin \chi \pi R}
\label{eq:fprop}
\end{equation}
is indeed the correct expression for the Feynman-propagator (with $\chi=\sqrt{p^2-m^2+i \epsilon}$). A different way to obtain this is via a basis of solutions to the Dirac-equation, followed by canonical quantization. To shed more light in the
construction of the propagator this possibility is explored in Appendix (\ref{sec:s1_canquant}).

\subsection{Gauge boson propagator}

Before writing down the Lagrangian one should make a convenient choice of gauge fixing. Since compactification
breaks down 5D to 4D Lorentz-invariance there is no need to fix the gauge in a way that respects the 5D symmetry. Therefore
one can choose a gauge fixing that is 4D Lorentz-invariant and use the remaining freedom to make the $A_a^5$ decouple from
the $A_a^\mu$ field \cite{Ghilencea:2001bw}.
\begin{align}
&\mathcal{L}_{\mathrm{gauge}}&=& -\frac{1}{4} F_{A B}^a F^{A B}_a -\frac{1}{2 \xi} \big(\partial_\mu A^\mu_a + \xi \partial_5 A^5_a\big)^2\nonumber\\
&&=&- \frac{1}{2} \partial_\mu A_\nu^a \partial^\mu A^\nu_a + \frac{1}{2} \partial_5 A^a_\nu \partial_5 A^\nu_a +\frac{1}{2} (1-\frac{1}{\xi}) \big(\partial_\mu A^\mu_a\big)^2\nonumber\\
&&&\hspace{3cm}+\frac{1}{2} \partial_\mu A^5_a \partial^\mu A^5_a - \frac{\xi}{2}\big(\partial_5 A^5_a\big)^2 + (\mathrm{interaction\ terms})\ .
\end{align}
This leads to the equations of motion
\begin{eqnarray}
\big(\Box  -(\partial_5)^2\big) A_a^\mu -(1-\frac{1}{\xi})\partial^\mu \partial_\nu A_a^\nu &=&0\nonumber\\
\big(\Box  - \xi (\partial_5)^2\big) A_a^5 &=&0\ .\label{eq:aem}
\end{eqnarray}
Green's functions of the above operators, satisfying the requirement of periodicity are given by (with $\chi=\sqrt{k^2+i \epsilon}$ and as before $y,y'\in[0,2\pi R[$)
\begin{multline}
D_{\mu \nu}=\int\!\! \frac{\deriv^4 k}{(2\pi)^4} (-i) e^{-ik.(x-x')} \bigg[\bigg(g_{\mu\nu} -\frac{k_\mu k_\nu}{\chi^2}\bigg) \frac{\cos \chi(\pi R - |y-y'|)}{2 \chi
\sin\chi\pi R} \\+ \frac{k_\mu k_\nu}{\chi^2} \frac{\cos (\chi/\sqrt{\xi})(\pi R -|y-y'|)}{2 (\chi/\sqrt{\xi})
\sin(\chi/\sqrt{\xi})\pi R}\bigg]
\end{multline}
\begin{equation}
D_{5 5}=\int\!\!\frac{\deriv^4 k}{(2\pi)^4}e^{-ik.(x-x')} \frac{i \cos(\chi/\sqrt{\xi})(\pi R - |y-y'|)}{2 (\chi/\sqrt{\xi})\sin(\chi/\sqrt{\xi})\pi R}\ , \qquad\quad D_{\mu 5}=0\ .
\end{equation}

\section{1-loop corrections on $\mathbb{R}^4 \times S_1$}
\label{sec:1ls1}

Although mixed momentum-/position-space propagators have been used in the context of extra dimensions at various
occasions (see for example \cite{Arkani-Hamed:1999za,Gherghetta:2000kr}), they have hardly ever been utilized to carry out loop calculations.
Generally going the traditional way by summing over KK-modes propagating in the loop always seemed to be easier.
There is however a special case where a number of authors have applied the method described here to do 1-loop calculations.
When the interaction is restricted to a fixed point (e.g. branes), the $y$-integration over the
interaction point collapses and is therefore trivial\footnote{This is for example the case in a number of supersymmetric theories, where supersymmetry forbids a Yukawa interaction. On the branes supersymmetry is partially broken and a Yukawa
interaction is allowed. Therefore, when calculating the correction to the Higgs-mass from a fermion loop, one only has to
deal with propagators that start and end at the brane.} . Examples of such calculations can be found in \cite{Arkani-Hamed:2001mi,Barbieri:2002uk}.\\
However we think that even in the case of bulk+brane interactions this method has its advantages. Especially when one calculates diagrams with only external zero-modes, which
are basically all diagrams needed for Standard-Model precision tests, evaluating the integrals over the extra dimension in
position space is generally rather simple.\\
For now we want to illustrate how to calculate radiative corrections using the propagators derived in the preceding section.
We will proceed by calculating the 1-loop corrections to the masses of the photon's KK-tower (for a treatment of this problem with
the conventional method see \cite{Cheng:2002iz}).

\subsection{Corrections to the KK-photon masses}

\label{sec:1loopphot}
\FIGURE[ht]{
\epsfig{file=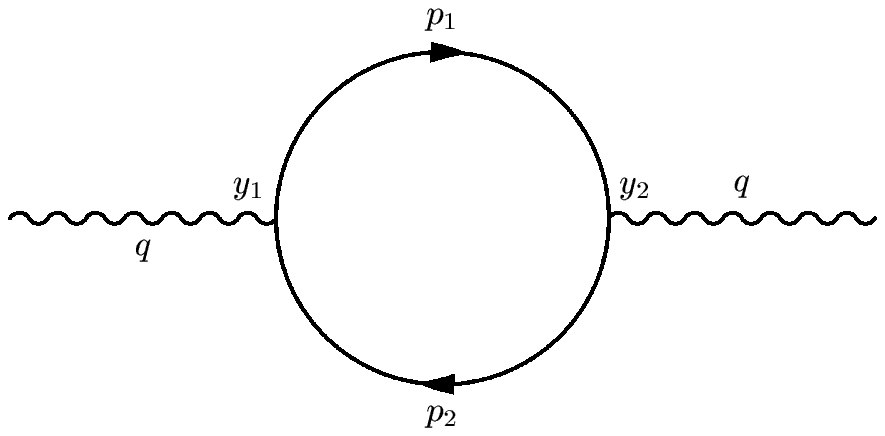}
\caption{Correction to the photon propagator}
\label{fig:kkmass}}
The 5D-QED Lagrangian with gauge fixing is given by
\begin{equation}
\label{eq:lag}
\mathcal{L}= -\frac{1}{4} F_{A B} F^{A B} - \frac{1}{2 \xi} (\partial_\mu A^\mu + \xi \partial_5 A^5)^2
 + i \bar\psi \gamma^A \mathcal{D}_A \psi\ .
\end{equation}
For reasons of simplicity we have set the 5D fermion mass to zero. As long as $m R \ll 1$ the corrections due to
a nonzero fermion mass will be negligible compared to those coming from the compactification. We have already
calculated the propagators for this theory in the preceding sections. What one still needs are the
wavefunctions of the free 1-particle photon-states. Solutions to Eq.(\ref{eq:aem}) normalized on the circle are
\begin{align}
&\mathrm{Zero-mode:}&&A^B_{0,+}(q,y)=\frac{1}{\sqrt{2\pi R}} \tilde{A}^B_{0,+}(q) \nonumber\\
&\mathrm{KK-modes:}&&A^B_{n,+}(q,y)=\frac{1}{\sqrt{\pi R}} \tilde{A}^B_{n,+}(q) \cos\frac{n y}{R}\nonumber\\
&&&A^B_{n,-}(q,y)=\frac{1}{\sqrt{\pi R}} \tilde{A}^B_{n,-}(q) \sin\frac{n y}{R} \ .\label{eq:phot_sol}
\end{align}
On-shell the coefficients $\tilde{A}^A_{n,\pm}$ fulfill the 4D wave equation of a massless/massive vector field
\begin{eqnarray}
q^2 \tilde{A}^\mu_{n,\pm} -\bigg(1-\frac{1}{\xi}\bigg) q^\nu q_\mu \tilde{A}^\mu_{n,\pm}-\frac{n^2}{R^2} \tilde{A}^\nu_{n,\pm} &=&0\nonumber\\
q^2 \tilde{A}^5_{n,\pm} - \xi \frac{n^2}{R^2} \tilde{A}^5_{n,\pm} &=&0\ .
\end{eqnarray}
Note that the KK-modes of $A^5$ can be gauged away by a suitably chosen gauge transformation and are eaten up by the
$A^\mu$ to form massive vector bosons. However the massless mode $A^5_0$ remains.
To obtain the mass shifts of the KK-states we need to calculate the photon self-energy.\\
A word concerning regularization: To make the integrals well defined we would, in principle, have to choose a regularization
method. However as we will find later (and also give a general argument for) the result will be finite and no renormalization
is needed. For reasons of simplicity we will therefore continue without regularizing the integrals that appear.\\
In QED there is only one diagram
(Fig.(\ref{fig:kkmass})), which is given by
\begin{multline}
\label{eq:1lphotmass}
\tilde{A}_{n,\pm}^{A} \big(i\,\Pi_{A B}^{n,\pm}\big) \tilde{A}_{n,\pm}^B = e^2\!\! \int\!\!\deriv^4 x_1\, \deriv^4 x_2\,
\deriv y_1\, \deriv y_2\,e^{-i p\cdot (x_2-x_1)}\times\\ A^{ A}_{n,\pm}(q,y_2) A^{B}_{n,\pm}(q,y_1)\tr \big[\gamma_A S_F(x_1-x_2,y_1,y_2) \gamma_B S_F(x_2-x_1,y_2,y_1)\big]\ .
\end{multline}
Since the manifold is translational invariant in the 5-direction $(\,\mathrm{mod}\, 2\pi R)$ momentum  in this
direction is conserved, making $\Pi_{\mu\nu}$ diagonal in the KK-label $n$. Also the mixing between the $+$ and $-$ states vanishes. Plugging
in the wavefunctions from Eq.(\ref{eq:phot_sol})
\begin{multline}
i\,\Pi_{A B}^{n,\pm} = \frac{-e^2}{(1+\delta_{n,0})2\pi R}\int\!\!\deriv y_1\,\deriv y_2\int\!\!\! \frac{\deriv^4 p}{(2\pi)^4} \bigg(\cos \frac{n(y_1-y_2)}{R} \pm\cos \frac{n(y_1+y_2)}{R}\bigg)\times\\
\tr\bigg[\gamma_A \bigg(\big(p_1^\sigma \gamma_\sigma +i \gamma^5\partial_{y_1}\big) \frac{\cos \chi_1(\pi R-|y_1-y_2|)}{2 \chi_1 \sin \chi_1 \pi R} \bigg)\times \\ \gamma_B \bigg(\big(p_2^\sigma \gamma_\sigma +i \gamma^5\partial_{y_2}\big) \frac{\cos \chi_2(\pi R-|y_2-y_1|)}{2 \chi_2 \sin \chi_2\pi R}\bigg)\bigg]\ .
\label{eq:se}
\end{multline}
Due to 4D momentum conservation
\begin{displaymath}
p_1=p,\quad p_2=p-q,\quad \chi_1=\sqrt{p^2+i\epsilon},\quad\chi_2=\sqrt{(p-q)^2+i \epsilon}\ .
\end{displaymath}
In a straightforward manner one can now evaluate the $y_1$,$y_2$ integrals. The result for the 4D part is given by
\begin{multline}
i\,\Pi^{\mu \nu}_{n,\pm} = 2 e^2\int\!\!\! \frac{\deriv^4 p}{(2\pi)^4}
\bigg[\frac{\cot \chi_1 \pi R}{\chi_1}\times\\
\hspace{2cm}\frac{\chi_1^2\big(\chi_1^2-\chi^2_2-\frac{n^2}{R^2}\big)g^{\mu\nu}+\big(\chi_1^2-
\chi_2^2+\frac{n^2}{R^2}\big)(p_1^\mu p_2^\nu+p_2^\mu p_1^\nu -g^{\mu\nu} p_1\cdot p_2)}{\big( (\chi_1-\chi_2)^2 - \frac{n^2}{R^2}\big)\big((\chi_1+\chi_2)^2-\frac{n^2}{R^2}\big)}+ (1\leftrightarrow 2)\bigg]\ ,
\label{eq:pse2}
\end{multline}
$\Pi^{\mu 5}=0$ and the $(5,5)$-component is
\begin{equation}
i\,\Pi^{5 5}_{n,\pm} = 2 e^2\int\!\!\! \frac{\deriv^4 p}{(2\pi)^4}
\bigg[\frac{\cot \chi_1 \pi R}{\chi_1}\frac{\chi_1^2\big(\chi_1^2-\chi^2_2-\frac{n^2}{R^2}\big)+\big(\chi_1^2-
\chi_2^2+\frac{n^2}{R^2}\big) p_1\cdot p_2}{\big( (\chi_1-\chi_2)^2 - \frac{n^2}{R^2}\big)\big((\chi_1+\chi_2)^2-\frac{n^2}{R^2}\big)}+ (1\leftrightarrow 2)\bigg]\ .
\label{eq:p5}
\end{equation}

As already discussed in Sec.(\ref{sec:S1_scal_prop}) this result consists of a finite term coming from the compactification
and a potentially divergent term that is the same as in the uncompactified theory. To identify the divergent part it is
instructive to perform the same calculation in a theory with five infinite dimensions. On finds
\begin{equation}
i\Pi^{A B}=e^2\int\!\!\frac{\deriv^5 p}{(2\pi)^5} \tr\left[\gamma^A \frac{i{p_1\!\!\!\!\!/}}{p_1^2+i\epsilon}\gamma^B 
\frac{i {p_2\!\!\!\!\!/}}{p_2^2 +i\epsilon}\right]\ ,
\end{equation}
which evaluates to
\begin{multline}
i\Pi^{\mu\nu}
=-e^2 \int\!\!\frac{\deriv^5 p}{(2\pi)^5} \frac{4(p_1^\mu p_2^\nu+p_2^\mu p_1^\nu-g^{\mu\nu} p_1\cdot p_2)+4(p_1^5 p_2^5)g^{\mu\nu}}{(p_1^2+i\epsilon)(p_2^2+i\epsilon)}
=2 e^2\int\!\!\frac{\deriv^4 p}{(2\pi)^4}\bigg[\frac{(-i)}{\chi_1}\times\\\qquad
\frac{\chi_1^2\big(\chi_1^2-\chi_2^2-(q^5)^2\big)g^{\mu\nu}+\big(\chi_1^2-\chi_2^2+(q^5)^2\big)(p_1^\mu p_2^\nu+p_2^\mu p_1^\nu-g^{\mu\nu} p_1
\cdot p_2)}{\chi_1\big(\chi_1-\chi_2)^2-(q^5)^2\big)\big(\chi_1+\chi_2)^2-(q^5)^2\big)}+(1\leftrightarrow 2)\bigg]
\label{eq:5dgaugemass}
\end{multline}
\begin{equation}
i\Pi^{5 5}=2 e^2\int\!\!\frac{\deriv^4 p}{(2\pi)^4}\bigg[\frac{(-i)}{\chi_1} \frac{\chi_1^2\big(\chi_1^2-\chi_2^2-(q^5)^2\big)+\big(\chi_1^2-\chi_2^2+(q^5)^2\big)p_1
\cdot p_2}{\chi_1\big(\chi_1-\chi_2)^2-(q^5)^2\big)\big(\chi_1+\chi_2)^2-(q^5)^2\big)}+(1\leftrightarrow 2)\bigg]\ .
\end{equation}
Using, that because of the small positive imaginary part of $\chi$,
\begin{equation}
\lim_{R\rightarrow \infty} \cot \chi \pi R =-i\qquad\mathrm{since}\quad \cot \chi \pi R= -i \Big(1+\frac{2}{e^{-2 i\pi R \chi}-1}\Big)
\end{equation}
one sees that this is just the $R\rightarrow\infty$ limit of the above equation (Eq.(\ref{eq:pse2})) with the 5-momentum $|q^5|=n/R$ held fixed.\\
To be precise one has to be a bit more careful. The external photon states defined in Eq.(\ref{eq:phot_sol}) correspond to a superposition
of states with positive/negative 5-momentum $q^5$. The $R\rightarrow\infty$ limit of $\Pi^{A B}_{n}$ on the
compactified manifold will therefore be the average of $\Pi^{A B}$ with $q^5=\pm n/R$ on the uncompactified manifold. In the
case presented here this is not important since Eq.(\ref{eq:5dgaugemass}) does not depend on the sign of $q^5$. It will however
be of relevance in the next section when we calculate the 1-loop correction to the fermion propagator.\\
Plugging the decomposition of the cotangent in Eq.(\ref{eq:pse2}) we now set forth to calculate the finite term
containing the exponential cutoff factor\footnote{This is in total analogy to
what one finds in finite temperature field theory. See \cite{Das:1997gg} for a calculation of the photon mass at
$T>0$.}.\\Note that in this particular case the contribution from the uncompactified theory Eq.(\ref{eq:5dgaugemass}) is
exactly zero. This is due to the fact that as $R\rightarrow\infty$ the full 5D Lorentz- and gauge-symmetry are restored.
Our result will therefore be manifestly finite, thus justifying our negligence of introducing a regulator. In what follows
we can make the replacement
\begin{equation}
\cot \chi\pi R\rightarrow \frac{-2 i}{e^{-i \chi (2\pi R)}-1}\ .
\label{eq:rep_cot}
\end{equation}
without changing any of the results. 
Calculating the masses of the KK-photons we want to take the on-shell limit of Eq.(\ref{eq:pse2}).
For $n\geq 1$ there is no problem letting $q^2\rightarrow n^2/R^2$. The case of the zero mode, where the denominator of Eq.(\ref{eq:pse2}) vanishes as $q^2\rightarrow 0$, will be treated later.
The remaining 4D Lorentz invariance requires $\Pi_{\mu\nu}$ to be of the form
\begin{equation}
\Pi_{\mu\nu}=g_{\mu\nu} \Pi_1 +q_\mu q_\nu \Pi_2\ .
\end{equation}
To calculate the mass corrections one is interested in $\Pi_1$ which can be extracted by
\begin{equation}
\label{eq:pi1}
\Pi_1 = \frac{1}{3}\Big(\Pi^\mu_{\ \mu}-\frac{1}{q^2}q^\mu q^\nu \Pi_{\mu \nu}\Big)\ .
\end{equation}
Plugging Eq.(\ref{eq:pse2}) in Eq.(\ref{eq:pi1}) and taking the on-shell limit yields (note that since the integral is
now convergent one can perform a shift in the integration variable of the second term of Eq.(\ref{eq:pse2}), yielding just twice the first term)
\begin{equation}
q^2 \Pi_1^{n}=\frac{4 e^2}{3}\int\!\!\! \frac{\deriv^4 p}{(2\pi)^4} \frac{\chi_1^2-\chi_2^2-2 \frac{n^2}{R^2}}{\chi_1}
\frac{1}{e^{-2 i\pi R \chi_1}-1}\ .
\end{equation}
Performing a Wick-rotation ($\chi_{1,2}\rightarrow i \chi_{1,2}$) and using that under spherical integration
\begin{equation}
\int\!\!\deriv \Omega (\chi_1^2-\chi_2^2) =\int\!\!\deriv\Omega (2 p.q-q^2)=-\!\int\!\!\deriv\Omega\, q^2\ ,
\end{equation}
One ends up with
\begin{equation}
\Pi_1^n = -\frac{e^2}{2\pi^2}\int_0^\infty\!\!\deriv \chi_1 \frac{\chi_1^2}{e^{2 \pi R \chi_1}-1}\quad\Rightarrow
\delta m^2_n = -\frac{e^2}{2 \pi R} \frac{\zeta(3)}{4 \pi^4 R^2}\ .
\end{equation}
With $e$ being expressed in terms of the 4D coupling constant $e_4=e/\sqrt{2 \pi R}$ this yields exactly the
KK-number independent mass shift found in \cite{Cheng:2002iz}.\\
We have not yet dealt with the zero mode. First confirm that in this case $\Pi_0^{\mu\nu}$ is transversal. Tracing Eq.(\ref{eq:pse2}) with $q$ yields
\begin{equation}
i \Pi^\mu_{\ \nu,0}q^\nu= e^2 \int\!\!\! \frac{\deriv^4 p}{(2\pi)^4}\bigg(\frac{\cot \pi R \chi_1}{\chi_1} k_1^\mu -
\frac{\cot \pi R \chi_2}{\chi_2} k_2^\mu\bigg)\ .
\end{equation}
If we split again in the part coming from the uncompactified theory and the rest, the first vanishes because of the
5D Ward-identity, whereas the latter is convergent and vanishes after a shift in the integration variable of the
second term. Due to 4D Lorentz invariance the tensor structure is required to be 
\begin{equation}
\Pi_0^{\mu\nu}=\Big(g^{\mu \nu} -\frac{q^\mu q^\nu}{q^2}\Big) \Pi_0^1\ .
\end{equation}
It is then sufficient to calculate the trace $\Pi_{\ \mu,0}^\mu$. From Eq.(\ref{eq:pse2}) we get (after Wick-rotation)
\begin{equation}
\Pi^\mu_{\ \mu,0} = -2 e^2\int\!\!\! \frac{\deriv^4 p_E}{(2\pi)^4}\bigg[
\frac{3 \chi_1^2-\chi_2^2+q^2}{\chi_1(\chi_1^2-\chi_2^2)}\frac{1}{e^{2\pi R \chi_1}-1}+ \frac{3 \chi_2^2-\chi_1^2+q^2}{\chi_2(\chi_2^2-\chi_1^2)}\frac{1}{e^{2\pi R \chi_2}-1}\bigg]\ .
\end{equation}
Taking the $q^2\rightarrow 0$ limit and using that $\chi_2 \rightarrow \chi_1$
\begin{equation}
\lim_{q^2\rightarrow 0} \Pi^\mu_{\ \mu,0}= -2 e^2 \int\!\! \frac{\deriv^4 p_E}{(2\pi)^4}\frac{1}{\chi_1^3}\, \frac{\deriv}{\deriv \chi_1}\bigg[\frac{\chi_1^3}{e^{2\pi R \chi_1}-1}\bigg]=\frac{-e^2}{(2\pi)^2}
\int_0^\infty\!\! \deriv \chi_1\, \frac{\deriv}{\deriv \chi_1}\bigg[\frac{\chi_1^3}{e^{2\pi R \chi_1}-1}\bigg]=0
\end{equation}
Thus the zero mode of the photon does not receive a correction and stays massless. This is very important because
if there was such a correction this would imply that 4D gauge invariance of the low energy theory (the theory of zero-modes only) was broken. For the zero mode of $A_5$ we get from Eq.(\ref{eq:p5})
\begin{equation}
\Pi^{5 5}_{0} = - 2 e^2\int\!\!\! \frac{\deriv^4 p_E}{(2\pi)^4}\bigg[
\frac{3 \chi_1^2+\chi_2^2-q^2}{\chi_1(\chi_1^2-\chi_2^2)}\frac{1}{e^{2\pi R \chi_1}-1}+ \frac{3 \chi_2^2+\chi_1^2-q^2}{\chi_2(\chi_2^2-\chi_1^2)}\frac{1}{e^{2\pi R \chi_2}-1}\bigg]\ .
\end{equation}
\begin{eqnarray}
\delta m_{0,5}^2=-\lim_{q^2\rightarrow 0} \Pi^{5 5}_{0}&=& 4 e^2 \int\!\! \frac{\deriv^4 p_E}{(2\pi)^4}\, \frac{\deriv}{\deriv \chi_1}\bigg[\frac{\chi_1^3}{e^{2\pi R \chi_1}-1}\bigg]\nonumber\\&=&\frac{e^2}{2 \pi^2}
\int_0^\infty\!\!  \deriv \chi_1\, \chi_1^2\frac{\deriv}{\deriv \chi_1}\bigg[\frac{\chi_1^3}{e^{2\pi R \chi_1}-1}\bigg]
=-\frac{e^2}{2 \pi R}\frac{3\, \zeta(3)}{4 \pi^4 R^2}\ ,
\end{eqnarray}
which is in agreement with earlier calculations \cite{Cheng:2002iz,vonGersdorff:2002as,Hosotani:1983xw}. The higher
KK-modes of $A_5$ can be gauged away using 5D gauge freedom and therefore do not represent physical states.
We have seen that the above calculation, although carried out in a non-renormalizeable theory has yielded a manifestly
finite result. This feature is shared by all those corrections that would vanish in the uncompactified theory. Those
contributions that are due to the compactification contain exponential cutoff factors that render the corresponding integrals
finite. The scale of this exponential cutoff is determined by the compactification scale. This is obvious from a physical point
of view, because the physics at distances much smaller than the radius, is the same whether the extra dimension is
compactified or not.\\
Another point is that we have extended our momentum integrals up to arbitrary high momenta. This is not quite correct since
the theory, being an effective theory only, has a natural cutoff. However if this cutoff is much larger than $1/R$, its
effect on the momentum integrals will be small due to the exponential damping. If it is however comparable to the
compactification radius, the theory already becomes unpredictive at scales where the first new physics appears and is
therefore not very interesting altogether.

\subsection{Non-abelian case}

When going over to the non-abelian case, there are additional diagrams coming from vertices with three or four gauge-bosons.
The relevant terms in the Lagrangian are
\begin{multline}
\mathcal{L}_{\mathrm{gauge}}= - g f^{abc}\big( (\partial_\mu A_\nu^a) A^{\mu,b} A^{\nu,c}-
(\partial_\mu A^{5,a}) A^{\mu,b} A^{5,c}+(\partial_5 A_\mu^a) A^{5,b} A^{\mu,c}\big)\\-
\frac{1}{4} g^2 f^{abc} f^{ade}\big( A_\mu^b A_\nu^c A^{\mu,d} A^{\nu,e}- 2 A^{5,b} A_\mu^c A^{5,d} A^{\mu,e}\big) +(\mathrm{quadratic\ terms})\ .
\end{multline}
The gauge-fixing term becomes
\begin{equation}
\mathcal{L}_{g.f.}=-\frac{1}{2\xi} f^a f^a\ ,\qquad \mathrm{with}\ f^a=\partial_\mu A^{\mu,a} + \xi \partial_5 A^{5,a}\ .
\end{equation}
From this it follows that the ghosts are describes by
\begin{equation}
\mathcal{L}_{\mathrm{ghost}}=
-\bar{c}^a \big(\partial_\mu\partial^\mu - \xi (\partial_5)^2\big) c^a - g f^{abc}\bar{c}^a\big(
\partial_\mu A^{\mu,b} c^c + \partial_5 A^{5,b}c^c\big)\ ,
\end{equation}
and the propagator is given by
\begin{equation}
D_{\mathrm{ghost}} =\int\!\!\frac{\deriv^4 p}{(2\pi)^4} e^{-i k.(x-x')}\frac{i \cos(\chi/\sqrt{\xi}) (\pi R-|y-y'|)}{2 \chi/ \sqrt{\xi} \sin(\chi/\sqrt{\xi}) \pi R} \delta^{ab}\ ,\qquad \chi=\sqrt{k^2+i \epsilon}\ .
\end{equation}
Like in QED, in the limit $R\rightarrow \infty$, the gauge-boson-mass vanishes because of gauge-/Lorentz-invariance.
After compactification the remaining corrections are again finite. The correction to the mass of the zero-mode gauge boson vanishes as expected. Details of the calculation can be found in Appendix (\ref{sec:na_diag}).\\
For the KK-modes one finds a constant mass shift, independent of the KK-label
\begin{equation}
\delta m^2 = \frac{3}{4}\frac{g^2}{2 \pi R}\frac{\zeta(3)}{4\pi^2 R^2}
C_2(G)\ ,
\end{equation}
which coincides with the result obtained in \cite{Cheng:2002iz}.

\subsection{Corrections to KK-fermion mass}
\label{sec:fermmass}
\FIGURE[ht]{
\epsfig{file=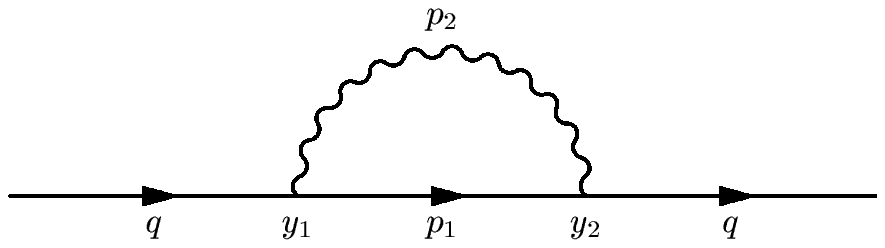}
\caption{Correction to the fermion propagator}
\label{fig:fermprop}
}
As a second example of a 1-loop calculation we evaluate the correction to the fermion mass, again in the
theory introduced
in the last section (Eq.(\ref{eq:lag})). As before the first thing one has to do is to find the wavefunctions of the
external states. Normalized on the circle they are given by
\begin{align}
&u_n^+ (p,y)=\frac{1}{\sqrt{(1+\delta_{n,0})\pi R}} \left(%
\begin{array}{cc}
  \cos \frac{n y}{R} & 0 \\
  0 & \phantom{-}\sin \frac{n y}{R} \\
\end{array}%
\right)\cdot \tilde{u}_\frac{n}{R} (p)\nonumber\\
&u_n^- (p,y)=\frac{1}{\sqrt{(1+\delta_{n,0})\pi R}} \left(%
\begin{array}{cc}
  \sin \frac{n y}{R} & 0 \\
  0 & -\cos \frac{n y}{R} \\
\end{array}%
\right)\cdot \tilde{u}_\frac{n}{R} (p)\nonumber\\
&v_n^+ (p,y)=\frac{1}{\sqrt{(1+\delta_{n,0})\pi R}} \left(%
\begin{array}{cc}
  \cos \frac{n y}{R} & 0 \\
  0 & -\sin \frac{n y}{R} \\
\end{array}%
\right)\cdot \tilde{v}_\frac{n}{R} (p)\nonumber\\
&v_n^- (p,y)=\frac{1}{\sqrt{(1+\delta_{n,0})\pi R}} \left(%
\begin{array}{cc}
  \sin \frac{n y}{R} & 0 \\
  0 & \phantom{-}\cos \frac{n y}{R} \\
\end{array}%
\right)\cdot \tilde{v}_\frac{n}{R} (p)\ .
\label{eq:psisol}
\end{align}
On-shell $\tilde{u}_\frac{n}{R}$,$\tilde{v}_\frac{n}{R}$ are pos.-/neg.-energy solutions to the 4-dimensional Dirac-equation with mass $m=n/R$.
\begin{equation}
\Big(p\!\!\!/-\frac{n}{R}\Big) \tilde{u}_\frac{n}{R}(p)=0\ ,\qquad \Big(p\!\!\!/+\frac{n}{R}\Big) \tilde{v}_\frac{n}{R}(p)=0
\end{equation}
The only diagram that contributes (Fig.(\ref{fig:fermprop})) is then given by
\begin{multline}
\bar{\tilde{u}}_\frac{n}{R}\big(-i \Sigma_2^{n,\pm}\big)\tilde{u}_\frac{n}{R}=- e^2 \int\!\!\deriv^4 x_1 \deriv^4 x_2
\deriv y_1 \deriv y_2\ e^{-i q\cdot(x_1-x_2)} \times\\
\bar{u}_n^\pm (q,y_2)\gamma^A S_F(x_2-x_1,y_2,y_1)\gamma^B u_n^\pm(q,y_1) D_{A B}(x_1-x_2,y_1,y_2)\ .
\end{multline}
with an equivalent expression for the negative energy solutions. Due to momentum conservation in the fifth dimension this is diagonal in the KK-index and as in the case of the photon propagator there is no mixing between the + and - states. Contrary
to the case of the KK-photon mass there is no symmetry that renders the fermion mass finite. Therefore the momentum
integrals have to be properly regularized (we leave the regularization implicit, but assume that it justifies the
manipulations in the rest of this section).\\ Using
Eq.(\ref{eq:psisol}) and choosing Feynman gauge $\xi=1$ for reasons of simplicity, one can evaluate the $y_1$, $y_2$
integrals and finds
\begin{multline}
-i \Sigma_2^{n,\pm} = \frac{3 e^2}{2}\int\!\!\frac{\deriv^4 p}{(2\pi )^4}
\Bigg[\frac{2\chi_1^2 \frac{\cot \pi R \chi_1}{\chi_1}-\big(\chi_1^2+\chi_2^2-\frac{n^2}{R^2}\big)\frac{\cot
\pi R \chi_2}{\chi_2}}{\big((\chi_1 -\chi_2)^2-\frac{n^2}{R^2}\big)\big((\chi_1 +\chi_2)^2-\frac{n^2}{R^2}\big)}\frac{n}{R} \\
-\bigg(\frac{\chi_1^2-\chi_2^2+\frac{n^2}{R^2}}{\big((\chi_1 -\chi_2)^2-\frac{n^2}{R^2}\big)\big( (\chi_1 +\chi_2)^2-\frac{n^2}{R^2}\big)} \frac{\cot \pi R \chi_1}{\chi_1} +(1\leftrightarrow 2)\bigg)p\!\!\!/ \Bigg]\ ,
\label{eq:sig2}
\end{multline}
again with
\begin{displaymath}
\chi_1=\sqrt{p^2+i \epsilon},\qquad \chi_2=\sqrt{(p-q)^2+i \epsilon}\ .
\end{displaymath}
For comparison it is useful to calculate the corresponding expression in the uncompactified theory.
\begin{eqnarray}
-i \Sigma_{2} &=& -e^2 \int\!\!\frac{\deriv^5 p}{(2\pi)^5}\gamma^A \frac{i({p\!\!\!/}_1 - p_1^5 \gamma^5)}{p_1^2+i\epsilon}
\gamma_A\frac{(-i)}{p_2^2+i\epsilon}\nonumber\\
&=&\frac{-3 i e^2}{2}\int\!\!\frac{\deriv^4 p}{(2\pi)^4}\frac{1}{(\chi_1+\chi_2)^2-(q^5)^2} \bigg[\frac{1}{\chi_1}{p\!\!\!/}_1 
+\frac{1}{\chi_2}\big({p\!\!\!/}_1 -q^5 \gamma^5\big)\bigg]
\label{eq:sig2uc}
\end{eqnarray}
This is yet not quite the $R\rightarrow\infty$ limit of Eq.(\ref{eq:sig2}). The reason for this is that, as discussed before, an external states is
 a superposition of states with 5-momentum $q_5=\pm n/R$. Eigenstates of $q_5$ are given by
\begin{align}
&u(q,q^5)=\left(%
\begin{array}{cc}
  1 & 0 \\
  0 & \pm i \\
\end{array}%
\right)\cdot \tilde{u}_{|q^5|} e^{-i q_5 y} &\mathrm{for}\ q_5\gtrless 0&\nonumber\\
&v(q,q^5)=\left(%
\begin{array}{cc}
  1 & 0 \\
  0 & \pm i \\
\end{array}%
\right)\cdot \tilde{v}_{|q^5|} e^{i q_5 y} & \mathrm{for}\ q_5\gtrless 0&
\end{align}
Averaging Eq.(\ref{eq:sig2uc}) over the propagation of a particle with pos./neg. 5-momentum gives (again the same is true
for the negative-energy solutions)
\begin{eqnarray}
\bar{\tilde{u}}_{|q^5|} \big(-i \Sigma'_{2}\big) \tilde{u}_{|q^5|}&=&\frac{1}{2}\bigg[u(q,q^5)\big(-i \Sigma_{2}\big)u(q,q^5)+
u(q,-q^5)\big(-i \Sigma_{2}\big)u(q,-q^5)\bigg]\nonumber\\
\Rightarrow -i \Sigma'_{2}&=&\frac{-3 i e^2}{2}\int\!\!\frac{\deriv^4 p}{(2\pi)^4}\frac{1}{(\chi_1+\chi_2)^2-(q^5)^2} \bigg[\frac{1}{\chi_1}{p\!\!\!/}_1 
+\frac{1}{\chi_2}\big({p\!\!\!/}_1 +|q^5|\big)\bigg]\ .
\end{eqnarray}
One can explicitly check that this is indeed the limit of Eq.(\ref{eq:sig2}) for infinite radius.\\
To calculate the correction to the fermion mass one needs to evaluate the propagator on-shell replacing $q\!\!\!/ \rightarrow
n/R$. After subtracting the $R\rightarrow\infty$ limit by replacing the cotangent by the exponential factor, one can
shift the integration variable and finds
\begin{multline}
-i \Sigma_{2,\mathrm{sub.}}^{n} = \frac{3 e^2}{2}\int\!\!\frac{\deriv^4 p}{(2\pi )^4}
\frac{\chi_1^2-\chi_2^2+\frac{n^2}{R^2}}{\big((\chi_1 -\chi_2)^2-\frac{n^2}{R^2}\big)\big((\chi_1 +\chi_2)^2-\frac{n^2}{R^2}\big)}\frac{-2 i}{e^{-i \chi_1 (2\pi R)}-1}\bigg[\frac{n}{R} -q\!\!\!/\bigg]
\end{multline}
From this one immediately sees that 
\begin{equation}
\delta \tilde{m}^n \simeq \Sigma_{2,\mathrm{sub.}}^{n}\big|_{q\!\!\!/\rightarrow \frac{n}{R}} =0
\end{equation}
which is in accordance with \cite{Cheng:2002iz}. Note however that the term coming from the uncompactified theory in this
case does not vanish. It is divergent and can only be evaluated in terms of an 
effective field theory when it will strongly depend on the cutoff.

\section{Propagators on $\mathbb{R}^4 \times S_1/Z_2$}
\label{sec:orb}
\subsection{Orbifold compactification}
In the preceding section we have dealt with compactifications on a circle. However there is a serious drawback that
makes such theories unsuitable as candidates for a physical theory. The problem is that in five dimensions the Dirac
spinor is an irreducible representation of the proper orthochronous Lorentz-group and not as in four dimensions a
reducible one. This implies that there is no Lorentz-invariant theory that is not left/right symmetric, which is
unsuitable for a theory of
weak interactions. One possible cure for this problem is the introduction of an additional discrete space-time symmetry.
The simplest example is the orbifold $S_1/Z_2$, where one introduces the reflection-symmetry $y\rightarrow 2\pi R -y$.
To promote this to a symmetry of the theory one has specify $Z_2$ parities for the various fields. Fields with even
parity will have low energy zero modes in addition to their KK-towers. For those of odd parity the zero modes are 
projected out. This means that in the low energy limit (much lower than the compactification radius), where the theory
is effectively a theory of zero-modes, the left/right symmetry can be lifted by  assigning different parities to the left-/right-handed components of the Dirac spinor.

\subsection{Scalar propagator}
We start with the same Lagrangian as in Sec.(\ref{sec:S1_scal_prop}) but in addition require that the
resulting theory is invariant under the action of $Z_2$. Introducing an operator $\mathcal{P}_5$ that implements
a reflection around $\pi R$ taking $y\rightarrow 2\pi R-y$, we impose the following conditions on the scalar
fields
\begin{equation}
\phi(x,y) = \mathcal{P}_5 \phi(x,y)=\wp_\phi \phi(x,2\pi R-y)\ .
\end{equation}
The sign depends on whether one assigns a positive or negative $Z_2$ parity $\wp_\phi =\pm 1$ to the field. One could now
use the same trick as in Eq.(\ref{eq:comprop}) to get the propagator on the Orbifold. In addition to applying the group of
translations by $2\pi R$ to the propagator and summing over it, one would also sum the contributions obtained by acting with $Z_2$. The physical interval would then be $y\in[0,\pi R]$. However to avoid boundary terms in the path-integral it is advantageous not to define the theory on an interval but to extend it to the full circle. Following an idea proposed in \cite{Georgi:2000ks} we define the new propagator to be
\begin{equation}
S_F^{\mathrm{orb.}} (x-x',y,y')= \frac{1}{2}\sum_{n=-\infty}^{+\infty} \big(S_F(x-x',y + 2 \pi R n,y') +\wp_\phi
S_F(x-x',2 \pi R n-y,y')\big)\ ,
\label{eq:orbsprop}
\end{equation}
which is exactly half of what one would get with the method described above. The integration in the path-integral now
extends over all of $S_1$. Note that the two ways of defining the theory are equivalent. It is just a matter of
convenience when doing calculations. For $y,y'\in [0,\pi R]$ one can evaluate Eq.(\ref{eq:orbsprop}) and finds
\begin{equation}
S_F^{\mathrm{orb.}} (x-x',y,y')=\int\!\!\!\frac{\deriv^{4} p}{(2\pi)^{4}}\frac{i \big(\cos \chi (\pi R-|y-y'|)
+\wp_\phi \cos \chi (\pi R-(y+y'))\big)}{4 \chi \sin \chi \pi R} e^{-i p\cdot (x-x')}\ .
\label{eq:scalorbprop}
\end{equation}
The second term under the integral gives an exponential cutoff factor except when both $y$ and $y'$ are on a "brane".
After Wick-rotation ($\chi=i \chi_E$) this term becomes
\begin{multline}
\frac{-i \wp_\phi \cosh \chi_E(\pi R -(y+y'))}{4 \chi_E \sinh \chi_E \pi R}\\\xrightarrow[p_E^2 \gg R^{-2}]\ \frac{-i \wp_\phi} {4\chi_E\, e^{\chi_E(\pi R -|\pi R-(y+y')|)}}\xrightarrow[p_E^2\rightarrow \infty] \
\Bigg\{ \begin{array}{c}
  0\ \ \mathrm{for}\ y,y'\!\in\,]0,\pi R[\ \vspace{1mm}\\
  \frac{-i \wp_\phi}{4\chi_E}\ \mathrm{for}\ y\!=\!y'\!\in\!\{0,\pi R\}\ .\\
\end{array}
\end{multline}
This tells us that in the case of the orbifold there will be additional divergencies that do not coincide with those of the
uncompactified theory. When calculating radiative corrections in Sec.(\ref{sec:1lorb}) we will see that the corresponding
counterterms are located on the branes \cite{Georgi:2000ks}. This is in accordance with the symmetries of our theory since at these points
local translation invariance is broken. Therefore in the case of the orbifold one has to allow for terms in the
Lagrangian that are located on the branes.\\
It is an easy task to extend the propagator (Eq.(\ref{eq:orbsprop})) to all of $S_1$. Simply apply the parity transformations $\mathcal{P}_5$ acting on either $y$, $y'$ or on both.
On $S_1$ the propagator fulfills the following modified equation for a Green's function (with appropriate boundary conditions)
\begin{equation}
\big(-\Box  + (\partial_5)^2 - m^2\big)S_F^{\mathrm{orb.}}(x-x',y,y')=\frac{i}{2}\big(\delta(y-y')+\wp_\phi\delta(2\pi R-(y+y'))
\big)\delta^{(4)}(x-x')\ .
\end{equation}

\subsection{Fermion propagator}
For the fermions the $Z_2$ transformation is implemented via
\begin{equation}
\mathcal{P}_5 \psi(x,y)=\wp_\psi (i\gamma^5)\psi(x,2\pi R-y)\ .
\end{equation}
This has the consequence that the upper(left-handed) component of the Dirac spinor has a different
$Z_2$ parity than the lower(right-handed) component. Implementing the symmetry on the theory by requiring that
\begin{equation}
\psi(x,y)=\mathcal{P}_5\psi(x,y)=\wp_\psi (i \gamma^5) \psi(x,2\pi R-y)
\end{equation}
therefore leads to a zero mode for $\psi$ that is left-/right-handed depending on whether $\wp_\psi=\pm1$.
The fermion propagator is derived using the same trick as in the scalar case.
There is however one subtle point. Since $m\bar\psi \psi$ transforms odd under $Z_2$ it is not an
allowed term of the Lagrangian anymore. There are several ways to write down a more complicated mass term
that is allowed by the $Z_2$-symmetry \cite{Delgado:2002xf}. For now we will simply stick to the
massless case.
The propagator can be obtained from Eq.(\ref{eq:fprop}) by
\begin{align}
&S_F^{\mathrm{orb.}}(x-x',y,y')&&= \frac{1}{2}\big(S_F^c(x-x',y,y')+\wp_\psi (i\gamma^5)S_F^c(x-x',2\pi R-y,y')\big)\nonumber\\
&&&=\int\!\!\frac{\deriv^{4} p}{(2\pi)^{4}}e^{-i p\cdot (x-x')} \big(p\!\!\!/+i \gamma^5\partial_{y}\big)\times\nonumber\\
&&&\hspace{1cm}\frac{i \big(\cos \chi (\pi R-|y-y'|)
-\wp_\psi (i\gamma^5) \cos \chi (\pi R-(y+y'))\big)}{4 \chi \sin \chi \pi R}\ .
\end{align}
Again this is valid only for $y,y'$ being on the interval $[0,\pi R]$. Extension of this result to the full circle
is as before obtained by acting with $\mathcal{P}_5$ from the left and/or from the right. The Green's function fulfills
\begin{equation}
\big(i\partial\!\!\!/+i \gamma^5\partial_y\big) S_F^{\mathrm{orb.}}(x-x',y,y')=
\frac{i}{2}\big(\delta(y-y')-\wp_\psi(i\gamma^5)
\delta(2\pi R-(y+y')\big)\delta^{(4)}(x-x')
\end{equation}

\subsection{Gauge-boson propagator}

The $Z_2$-parity assignment for the gauge fields is somewhat canonical if one requires that the covariant derivatives
transform appropriately. The $A_\mu$ fields are assigned an even parity, whereas the $A_5$ is assigned an odd parity.
\begin{align}
&A_\mu(x,y)=\mathcal{P}_5 A_\mu(x,y)=\phantom{-}A_\mu(x,2\pi R-y)\nonumber\\
&A_5(x,y)=\mathcal{P}_5 A_5(x,y)=-A_5(x,2\pi R-y)
\end{align}
The propagators take on the form (for $y,y'\in[0,\pi R]$)
\begin{multline}
D^{\mathrm{g.f}}_{\mu \nu}=\int\!\! \frac{\deriv^4 k}{(2\pi)^4} (-i)  \bigg(\bigg(g_{\mu\nu} -\frac{k_\mu k_\nu}{\chi^2}\bigg) \frac{\cos \chi(\pi R - |y-y'|)+\cos \chi(\pi R - (y+y'))}{4 \chi
\sin\chi\pi R} \\+ \frac{k_\mu k_\nu}{\chi^2} \frac{\cos (\chi/\sqrt{\xi})(\pi R -|y-y'|)+\cos (\chi/\sqrt{\xi})
(\pi R -(y+y'))}{4 (\chi/\sqrt{\xi})
\sin(\chi/\sqrt{\xi})\pi R}\bigg)e^{-ik.(x-x')}
\end{multline}
\begin{equation}
D^{\mathrm{g.f}}_{5 5}=\int\!\!\frac{\deriv^4 k}{(2\pi)^4}i \frac{\cos(\chi/\sqrt{\xi})(\pi R - |y-y'|)-
\cos(\chi/\sqrt{\xi})(\pi R - (y+y'))}{4 (\chi/\sqrt{\xi})\sin(\chi/\sqrt{\xi})\pi R}e^{-ik.(x-x')}
\end{equation}
and fulfill the following equations on $S_1$
\begin{align}
&\big(\Box  -(\partial_5)^2\big) D^\mu_{\ \nu} -\Big(1-\frac{1}{\xi}\Big)\partial^\mu \partial^\lambda D_{\lambda \nu}
 =\frac{i}{2}\delta^\mu_{\ \nu} \big(\delta(y-y')+\delta(2\pi R-(y+y'))\big)\delta^{(4)}(x-x')\nonumber\\
&\big(\Box  - \xi (\partial_5)^2\big) D^5_{\ 5} =\frac{i}{2}\big(\delta(y-y')-\delta(2\pi R-(y+y'))\big)\delta^{(4)}(x-x')
\ .\label{eq:aorbprop}
\end{align}

\section{1-loop corrections on $\mathbb{R}^4 \times S_1/Z_2$}
\label{sec:1lorb}

We now proceed by calculating the one-loop corrections evaluated in Sec.(\ref{sec:1ls1}) in the case of the orbifold. However
we have to face an additional subtlety. As already discussed in the previous section, there will be divergent contributions
that are located on the branes. These violate translational invariance and therefore momentum conservation and introduce
mixing between states with different KK-labels (momentum in the 5-direction).

\subsection{Corrections to KK-photon mass}
We start by calculating the 1-loop photon propagator corrections in 5D QED formulated on the orbifold (for the same calculation done with the conventional method see again \cite{Cheng:2002iz}). The Lagrangian is the same as in the case
of $S_1$, Eq.(\ref{eq:lag}). Because
of the $Z_2$ symmetry only the even photon states of Eq.(\ref{eq:phot_sol}) survive. Performing the same
calculation as before (Eq.(\ref{eq:1lphotmass})), one finds for $\Pi^{\mu\nu,\mathrm{orb.}}_{n,n'}$
\begin{equation}
\label{eq:1lorb}
\Pi^{\mu\nu,\mathrm{orb.}}_{n}=\frac{1}{2}\Pi^{\mu\nu}_{n}\ ,
\end{equation}
one half of the result of Sec.(\ref{sec:1loopphot}). This is due to the fact that the $Z_2$ symmetry has projected out half of the states that propagate in the fermion-loop. There is however another term that
is not diagonal in the KK-labels. It comes from traces like $\tr[\gamma^\mu \gamma^\nu\gamma^\rho\gamma^\lambda\gamma^5]$ and
is given by
\begin{equation}
\label{eq:boundterm}
\sim\frac{e^2}{\pi R}\int\!\!\frac{\deriv^4 p}{(2\pi)^4} \left(\frac{i\epsilon^{\mu\nu\rho\lambda}
k^1_\rho k^2_\lambda}{\left(\chi_1^2-\left(\frac{n-n'}{2 R}\right)^2\right)\left(\chi_2^2-\left(\frac{n+n'}{2 R}\right)^2\right)} - (1\leftrightarrow 2)\right)=0\ .
\end{equation}
Because of 4D Lorentz invariance, the only available Lorentz tensors are $g^{\mu\nu}$ and $q^\mu q^\nu$ which
are both symmetric in $\mu,\nu$ and therefore the above expression has to vanish.
Using the results from the previous sections the mass shift is exactly half of what it was in the case of $S_1$
\cite{Cheng:2002iz}
\begin{equation}
\delta m_n^2=-\frac{e^2}{2 \pi R} \frac{\zeta(3)}{8 \pi^4 R^2}\ .
\end{equation}

\subsection{Corrections to KK-fermion mass}
Carrying out the same steps as in Sec.(\ref{sec:fermmass}) for the case of the orbifold yields
\begin{equation}
\Sigma_{2,\mathrm{orb.}}^{n,n'}= \frac{1}{2}\Sigma_2^{n} \delta_{n,n'} + \Sigma_{2,\mathrm{brane}}^{n,n'}\ .
\end{equation}
The first term is again half of what we have found in the case of compactification on $S_1$, diagonal in 
the KK-indices and independent of the $Z_2$ parity chosen for the fermion field. However the second term
depends on the parity and introduces a mixing between modes of different KK-labels. This term is given by
\begin{align}
&\mathrm{for}\ n-n'\ \mathrm{odd:}\nonumber\\
&-i \Sigma_{2,\mathrm{brane}}^{n,n'}=0\nonumber\\
&\mathrm{for}\ n-n'\ \mathrm{even}:\nonumber\\
&-i \Sigma_{2,\mathrm{brane}}^{n,n'}=
\frac{e^2}{\sqrt{(1+\delta_{n,0})(1+\delta_{n',0})}8 \pi R} \int\!\!\frac{\deriv^4 p}{(2\pi)^4}\times\nonumber\\
&\qquad\Bigg[ \frac{2 {p\!\!\!/}_1 - 5 \frac{n+n'}{R}}{\left(\chi_1^2-
\left(\frac{n+n'}{2 R}\right)^2\right)\left(\chi_2^2-\left(\frac{n-n'}{2 R}\right)^2\right)}- \wp_\psi i\gamma^5
\frac{2 {p\!\!\!/}_1 - 5 \frac{n-n'}{R}}{\left(\chi_1^2-
\left(\frac{n-n'}{2 R}\right)^2\right)\left(\chi_2^2-\left(\frac{n+n'}{2 R}\right)^2\right)}\Bigg]\ .
\end{align}
It violates momentum conservation in the fifth direction (by multiples of $2/R$), and leads to a mixing of different KK-modes. Going to Euclidean momentum it is easy to extract the leading logarithmic singularity
\begin{equation}
\Sigma_{2,\mathrm{brane}}^{n,n'}\simeq
\frac{-e^2}{\sqrt{(1+\delta_{n,0})(1+\delta_{n',0})}64 \pi^3 R} \ln \frac{\Lambda}{\mu} 
\Big[ q\!\!\!/(1- \wp_\psi i\gamma^5) - 5 \frac{n}{R}(1- \wp_\psi i\gamma^5)-5 \frac{n'}{R}(1+\wp_\psi \gamma^5)\Big]\ .
\end{equation}
Since local 5D Lorentz-symmetry is broken by the orbifold, terms that are located on the fixed points are allowed by the
symmetries. The cancellation of the above divergencies requires the introduction of counterterms that live on the branes. These additional terms have to be added to the Lagrangian and even when finetuned to zero they will be generated
at loop-level. Also for the case of large mass terms located on the boundary, the Eigenstates would be shifted. We will
therefore assume that the tree-level boundary mass-term is sufficiently small not to significantly alter the picture
presented above. Radiative corrections to this value will then be loop suppressed.

\section{Conclusion}
Dealing with a theory formulated on a spacetime with compactified dimensions one has several options. The well beaten path
is to integrate out the additional dimensions and work with a 4D theory of infinitely many Kaluza-Klein modes. Here we have studied an alternative approach especially suitable to five dimensions, where one retains the additional dimension, but instead of working in momentum space as for the four "normal" dimensions, treats the additional dimension in configuration space. We have calculated the propagators for some example setups and have illustrated how to do 1-loop calculations in this formalism. Also we have demonstrated that the UV-divergencies of the
compactified and the corresponding uncompactified theory are the same and have explicitly shown how to extract those
contributions that are due to the compactification. These are in general finite except for the case of the orbifold, where
divergent terms that are located on the branes can arise.
Higher dimensional symmetries (Lorentz, gauge, ...) may require some physical observables to vanish in the uncompactified
theory. After compactification these quantities receive only finite corrections (except for possible boundary terms)
and can be calculated.\\
There are several advantages of working with the mixed momentum-/configuration-space formalism.
The derivation of the propagator is straightforward, since it is not necessary to find the mass Eigenstates that propagate. Especially in the case of diagrams with external zero-modes (Standard Model precision data) the integrals over the compactified dimension can be evaluated easily. Even in the case of external KK-modes the integrals can be done with the aid of some mathematical manipulation software in a straightforward fashion. Similarly to finite temperature field-theory the relation between the compactified ($T>0$) and the corresponding uncompactified ($T=0$) theory is more obvious. One disadvantage
is that in more than five dimensions many of the advantages (no summation over modes, simple form of the propagator) are lost.
\newpage
\appendix
\section{Canonical quantization of fermion fields}
\label{sec:s1_canquant}
The requirement of periodicity in the compactified dimension forces the solutions to the Dirac-equation to be a
superposition of
\begin{displaymath}
\sin \frac{n y}{R}\ ,\quad \cos\frac{n y}{R}\qquad \mathrm{with}\ p^2=m^2+\frac{n^2}{R^2}\ .
\end{displaymath}
From the mass-shell condition on sees that from the $4D$ point of view one gets the familiar picture of an infinite tower
of Kaluza-Klein states.
To fix a basis we introduce two sets of orthogonal and normalized 2-spinors $\xi^s$ and $\eta^s$ ($s \in \{1,2\}$) with $\xi^{s\dagger} \xi^{r} =\delta^{s,r}$ and $\eta^{s\dagger} \eta^{r} =\delta^{s,r}$. A complete orthonormal set
of positive energy solutions of the Dirac equation (Eq.(\ref{eq:s1_direq})) is then given by\footnote{The matrices $\sqrt{p_\mu \sigma^\mu}$ and $\sqrt{p_\mu \bar{\sigma}^\mu}$
can be completely fixed by requiring that $\sqrt{p_\mu \sigma^\mu}^2 = p_\mu \sigma^\mu$, $\sqrt{p_\mu \bar{\sigma}^\mu}^2 = p_\mu \bar{\sigma}^\mu$ and that $\sqrt{p_\mu \bar{\sigma}^\mu} \cdot \sqrt{p_\mu \sigma^\mu}= \sqrt{p_\mu \sigma^\mu}\cdot
\sqrt{p_\mu \bar{\sigma}^\mu} = \sqrt{p^2}$ (this construction of Dirac-spinors follows \cite{Peskin:1995ev}).}
\begin{align}
&u^{+}_{s,0}(p,y)=\left(%
\begin{array}{c}
   \sqrt{p\cdot\sigma} \xi^s\vspace{1mm}\\
   \sqrt{p\cdot\bar{\sigma}} \xi^s \\
\end{array}%
\right) \qquad \mathrm{with}\ \ p^2=m^2+\frac{n^2}{R^2} \nonumber\\
&u^{+}_{s,n}(p,y)=\sqrt{2}\left(%
\begin{array}{cc}
  \cos \frac{n y}{R} &\sqrt{p\cdot\sigma} \xi^s\vspace{1mm}\\
  \frac{m \cos\frac{n y}{R} + \frac{n}{R}\sin \frac{n y}{R}}{\sqrt{m^2+\frac{n^2}{R^2}}} &\sqrt{p\cdot
  \bar{\sigma}} \xi^s \\
\end{array}%
\right) \nonumber\\
&u^{-}_{s,n}(p,y)=\sqrt{2}\left(%
\begin{array}{cc}
  \sin \frac{n y}{R} &\sqrt{p\cdot\sigma} \xi^s\vspace{1mm}\\
  \frac{m \sin\frac{n y}{R} - \frac{n}{R}\cos \frac{n y}{R}}{\sqrt{m^2+\frac{n^2}{R^2}}} &\sqrt{p\cdot
  \bar{\sigma}} \xi^s \\
\end{array}%
\right) \ .
\end{align}
Note that for every higher KK-mode one gets two linearly independent solutions whereas for the zero mode one gets only one.
These solutions are normalized according to
\begin{equation}
\int\!\!\deriv^4 x\!\int_0^{2\pi R}\!\!\!\!\!\deriv y\ \bar{u}^{\sigma}_{s,n}(p,y)u^{\sigma'}_{s',n'}(p',y)
e^{-i (p'-p)\cdot x}
=2 m\,\delta^{s,s'}\delta^{\sigma,\sigma'} (2 \pi R) \delta^{n,n'} (2\pi)^4 \delta^{(4)}(p-p')\ ,
\end{equation}
with the spin-sums
\begin{align}
&\sum_s u^{+}_{s,0}(p,y) \bar{u}^{+}_{s,0}(p,y')=\big(p\!\!\!/ + m\big)\nonumber\\
&\sum_s \Big(u^{+}_{s,n}(p,y) \bar{u}^{+}_{s,n}(p,y')+
u^{-}_{s,n}(p,y) \bar{u}^{-}_{s,n}(p,y')\Big)=2 \big(p\!\!\!/ +i \gamma^5\partial_5 +m\big)\cos \frac{n(y-y')}{R}\ .
\end{align}
For the negative energy solutions one finds
\begin{align}
&v^{+}_{s,0}(p,y)=\left(%
\begin{array}{c}
   \sqrt{p\cdot\sigma}\, \eta^s\vspace{1mm}\\
   -\sqrt{p\cdot
  \bar{\sigma}}\, \eta^s\quad \\
\end{array}%
\right) \qquad \mathrm{with}\ \ p^2=m^2+\frac{n^2}{R^2} \nonumber\\
&v^{+}_{s,n}(p,y)=\sqrt{2}\left(%
\begin{array}{cc}
  \cos \frac{n y}{R} &\sqrt{p\cdot\sigma}\, \eta^s\vspace{1mm}\\
  -\frac{m \cos\frac{n y}{R} + \frac{n}{R}\sin \frac{n y}{R}}{\sqrt{m^2+\frac{n^2}{R^2}}} &\sqrt{p\cdot
  \bar{\sigma}}\, \eta^s \\
\end{array}%
\right)\nonumber\\
&v^{-}_{s,n}(p,y)=\sqrt{2}\left(%
\begin{array}{cc}
  \sin \frac{n y}{R} &\sqrt{p\cdot\sigma}\, \eta^s\vspace{1mm}\\
  \frac{-m \sin\frac{n y}{R} + \frac{n}{R}\cos \frac{n y}{R}}{\sqrt{m^2+\frac{n^2}{R^2}}} &\sqrt{p\cdot
  \bar{\sigma}}\, \eta^s \\
\end{array}%
\right) 
\end{align}
with normalization
\begin{equation}
\int\!\!\deriv^4 x\!\int_0^{2\pi R}\!\!\!\!\!\deriv y\ \bar{v}^{\sigma}_{s,n}(p,y)v^{\sigma'}_{s',n'}(p',y)
e^{i (p'-p)\cdot x}=-2 m\,\delta^{s,s'}\delta^{\sigma,\sigma'} (2 \pi R) \delta^{n,n'} (2\pi)^4 \delta^{(4)}(p-p')\ .
\end{equation}
and spin sums
\begin{align}
&\sum_s v^{+}_{s,0}(p,y) \bar{v}^{+}_{s,0}(p,y')=\big(p\!\!\!/ -m\big)\nonumber\\
&\sum_s \Big(v^{+}_{s,n}(p,y) \bar{v}^{+}_{s,n}(p,y')+
v^{-}_{s,n}(p,y) \bar{v}^{-}_{s,n}(p,y')\Big)=2 \big(p\!\!\!/ - i \gamma^5 \partial_y -m\big)\cos\frac{n(y-y')}{R}\ .
\end{align}
Proceeding with the usual construction of the Fock-space we introduce creation- and annihilation-operators with the
following anti-commutation relations
\begin{equation}
\{a^{\sigma,n}_{p,s}, a^{\sigma',n' \dagger}_{p',s'}\} =
\{b^{\sigma,n}_{p,s}, b^{\sigma',n' \dagger}_{p',s'}\} =(2\pi)^3 \delta^{(3)}(p-p')\,(2 \pi R)\,
\delta^{n,n'}\, \delta^{s,s'} \delta^{\sigma,\sigma'}
\end{equation}
and define 1-particle states
\begin{equation}
|p,n,s,\sigma\rangle = \sqrt{2 E_{p,n}}\ a^{\sigma,n \dagger}_{p,s}|0\rangle\ .
\end{equation}
These states are normalized according to
\begin{equation}
\langle p,n,s,\sigma|p',n',s',\sigma'\rangle = 2 E_{p,n} \,(2\pi)^3 \delta^{(3)}(p-p')\,(2 \pi R)\,
\delta^{n,n'}\, \delta^{s,s'} \delta^{\sigma,\sigma'}\ .
\end{equation}
This allows one to write down the quantized fields in terms of creation-/annihilation-operators
(note that $\int\frac{\deriv p^5}{2 \pi} \rightarrow \frac{1}{2 \pi R}\sum_{n=0}^\infty$)
\begin{align}
&\psi(x,y)=\frac{1}{2\pi R} \sum_{n=0}^\infty \int \frac{\deriv^3 p}{(2\pi)^3} \frac{1}{\sqrt{2 E_{p,n}}}
\sum_{s,\sigma} \Big(a^{\sigma,n}_{p,s}\, u^{\sigma}_{s,n}(p,y)e^{-i p\cdot x}+ b^{\sigma,n\dagger}_{p,s}\, v^{\sigma}_{s,n}(p,y)e^{i p\cdot x}\Big)\nonumber\\
&\bar{\psi}(x,y)=\frac{1}{2\pi R} \sum_{n=0}^\infty \int \frac{\deriv^3 p}{(2\pi)^3} \frac{1}{\sqrt{2 E_{p,n}}}\sum_{s,\sigma} \Big(b^{\sigma,n}_{p,s}\, \bar{v}^{\sigma}_{s,n}(p,y)
e^{-i p\cdot x} +a^{\sigma,n\dagger}_{p,s}\,\bar{u}^{\sigma}_{s,n}(p,y)e^{i p\cdot x}\Big)\ .
\end{align}
This allows one to calculate the 2-point Green's functions.
\begin{align}
&\langle 0| \psi_a(x,y) \bar{\psi}_b(x',y')|0\rangle=
\frac{1}{2\pi R} \sum_{n=0}^\infty \int\!\! \frac{\deriv^3 p}{(2\pi)^3} \frac{1}{2 E_{p,n}} 
\sum_{s,\sigma} u^{\sigma}_{s,n}(p,y) \bar{u}^{\sigma}_{s,n}(p,y')
e^{-i p\cdot(x-x')}\nonumber\\
&\hspace{1.5cm}=\frac{1}{\pi R} \sum_{n=0}^\infty \big(i \partial\!\!\!/ + i\gamma^5 \partial_y +m\big)_{a b}
\bigg[\cos\frac{n(y-y')}{R}-\frac{\delta_{n,0}}{2}\bigg] \int\!\! \frac{\deriv^3 p}{(2\pi)^3} \frac{1}{2 E_{p,n}}
e^{-i p\cdot(x-x')}\nonumber\\
&\langle 0| \bar{\psi}_b(x',y') \psi_a(x,y)|0\rangle =
\frac{1}{2\pi R} \sum_{n=0}^\infty \int\!\! \frac{\deriv^3 p}{(2\pi)^3} \frac{1}{2 E_{p,n}} 
\sum_{s,\sigma} v^{\sigma}_{s,n}(p,y) \bar{v}^{\sigma}_{s,n}(p,y')e^{-i p\cdot(x'-x)}\nonumber\\
&\hspace{1.5cm}=-\frac{1}{\pi R} \sum_{n=0}^\infty \big(i \partial\!\!\!/ + i\gamma^5 \partial_y +m\big)_{a b}
\bigg[\cos\frac{n(y-y')}{R}-\frac{\delta_{n,0}}{2}\bigg] \int\!\! \frac{\deriv^3 p}{(2\pi)^3} \frac{1}{2 E_{p,n}}
e^{-i p\cdot(x'-x)}
\end{align}
The time-ordered product of the two is given by
\begin{multline}
\langle 0| \mathsf{T}[\psi(x,y) \bar{\psi}(x',y')]|0\rangle=
\big(i\partial\!\!\!/_\mu + i\gamma^5 \partial_y +m\big)\times\\
\frac{1}{\pi R} \sum_{n=0}^\infty  \bigg[\cos\frac{n(y-y')}{R}-\frac{\delta_{n,0}}{2}\bigg] \int\!\! \frac{\deriv^4 p}{(2\pi)^4} \frac{i}{p^2-m^2+\frac{n^2}{R^2}
+i \epsilon}
e^{-i p\cdot (x-y)}\ .\label{eq:prop1}
\end{multline}
The sum can be evaluated for $y,y' \in [0,2 \pi R[$ making use of the following relation\footnote{See for example \cite{Magnus:1948}.} (for $x \in [2\pi m,2\pi (m+1)]$ and $a\notin \mathbb{Z}$)
\begin{align*}
&\sum_{n=1}^\infty \frac{\cos n x}{n^2-a^2}=\frac{1}{2 a^2}-\frac{\pi \cos a((2m+1)\pi-x)}{2 a \sin a \pi}
\end{align*}
and we find that (with $\chi = \sqrt{p^2-m^2+i \epsilon}$)
\begin{equation}
S^c_F(x-x',y,y')=\int\!\! \frac{\deriv^4 p}{(2\pi)^4} e^{-ip.(x-x')}
\big(p\!\!\!/ +i \gamma^5\partial_y + m\big) \frac{i \cos \chi(\pi R-|y-y'|)}{2 \chi \sin \pi R}\ ,
\end{equation}
confirming the result found earlier.
One can also check that this solution is indeed $i$ times the Green's function of the Dirac
operator
\begin{displaymath}
(\partial\!\!\!/ +i \gamma^5\partial_y - m\big) S_F(x-x',y,y')=i \delta^{(4)}(x-x') \delta(y-y')
\end{displaymath}
Generally the most efficient way to obtain the Feynman-propagator is to simply "guess" its form from the
corresponding expression for the scalar field. However it is instructive to see how the familiar procedure
of canonical quantization generalizes to the case of compactified extra dimensions.

\section{1-loop corrections to the non-abelian gauge boson propagator}
\label{sec:na_diag}
\vspace{5mm}
\FIGURE[ht]{
\epsfig{file=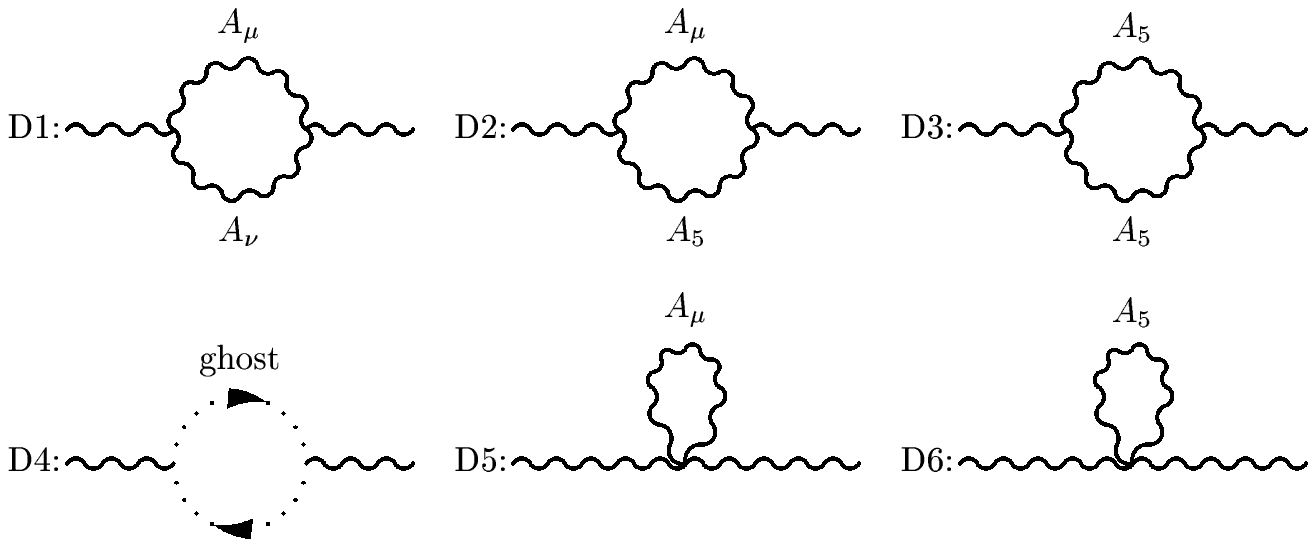}
\label{fig:diags}
\caption{Diagrams from the gauge sector contributing to $\Pi_2^{\mu\nu}$.}}
To calculate the correction to the masses of the gluon KK-tower one has to take into account the contributions coming
from fermion-loops (this is the same as in QED except for a factor of $T(r_f)=\tr[T^a,T^b]$) and from the gauge sector
(this is given by the diagrams in Table(\ref{fig:diags})).
Starting with diagram $D1$ one evaluates the $y$-integrals and finds that it is of the form
\begin{multline}
i\,\Pi^{\mu \nu}_{n,D1} = \frac{g^2}{4}\int\!\!\! \frac{\deriv^4 p}{(2\pi)^4}
\bigg[\frac{\cot \chi_1 \pi R}{\chi_1}\frac{\big(\chi_1^2-\chi^2_2+\frac{n^2}{R^2}\big)C_2(G) N^{\mu \nu}\delta^{ab}}{\big( (\chi_1-\chi_2)^2 - \frac{n^2}{R^2}\big)\big((\chi_1+\chi_2)^2-\frac{n^2}{R^2}\big)}+ (1\leftrightarrow 2)\bigg]\ ,
\end{multline}
where $N^{\mu\nu}$ is given by
\begin{multline}
N^{\mu\nu}=\big( (p_1+p_2)^\mu g^{\lambda\rho} - (p_1+q)^\lambda g^{\mu\rho}+(q-p_2)^\rho g^{\mu\lambda}\big)\times\\\big(
(-(p_1+p_2)^\nu g_{\lambda\rho}+(q+p_1)_\lambda \delta^\nu_{\ \rho} + (p_2-q)_\rho \delta^\nu_{\ \lambda}\big)\ .
\end{multline}
As in the case of the photon, the full $5D$ gauge- and Lorentz-symmetry again assures that those terms that are unaffected
by the $R\rightarrow \infty$ limit cancel. It is therefore legitimate to make the substitution Eq.(\ref{eq:rep_cot}). For $n>0$
the on-shell limit of the integrand is well defined and one gets a finite correction to the KK-mass given by
\begin{multline}
\delta \tilde{m}^{2}_{n,D1} = g^2 C_2(G)\int\!\!\! \frac{\deriv^4 p}{(2\pi)^4}
\bigg[\frac{1}{e^{-i 2 \pi R \chi_1}-1}\Big[-\frac{5}{6}\frac{\chi_1^2-\chi^2_2+\frac{n^2}{R^2}}{\chi_1\frac{n^2}{R^2}}\\+
\frac{\big(\chi_1^2-\chi^2_2+\frac{n^2}{R^2}\big)\big(\chi_1^2+\chi^2_2+4\frac{n^2}{R^2}\big)}{ \chi_1\big( (\chi_1-\chi_2)^2 - \frac{n^2}{R^2}\big)\big((\chi_1+\chi_2)^2-\frac{n^2}{R^2}\big)}\bigg]=\frac{1}{4}\frac{g^2}{2 \pi R}\frac{\zeta(3)}{4 \pi^4 R^2} C_2(G)\ .
\end{multline}
Here we have used that $C_2(G)\delta_{ab}=f^{acd}f^{bcd}$.
In the case of the zero-mode one once again has to be careful since the two terms in the integrand are divergent as
$q^2\rightarrow 0$. Nevertheless when taking the limit in a sensible way one gets
\begin{equation}
\delta \tilde{m}^{2}_{0,D1} =\! -\frac{9}{16}\frac{g^2}{\pi^2} C_2(G)\int_0^\infty\!\!\! \deriv \chi_1 \Bigg[
\frac{4 \chi_1^2}{e^{2 \pi R \chi_1}-1}-\frac{1}{4}\frac{\deriv}{\deriv \chi_1}\bigg[\frac{\chi_1^3}{e^{2 \pi R \chi_1}-1}\bigg]\Bigg]\!=\!-\frac{9}{8}\frac{g^2}{2 \pi R}\frac{\zeta(3)}{4 \pi^4 R^2} C_2(G)\,.
\end{equation}
For diagram $D2$ one finds
\begin{multline}
i\,\Pi^{\mu \nu}_{n,D2} = \frac{g^2}{2} C_2(G)\delta^{ab}\int\!\!\! \frac{\deriv^4 p}{(2\pi)^4}
\bigg[\frac{\cot \chi_1 \pi R}{\chi_1}\\+\frac{\cot \chi_1 \pi R}{\chi_1}\frac{\big(\chi_1^2-\chi^2_2-3\frac{n^2}{R^2}\big)\big(\chi_2^2-\frac{n^2}{R^2}\big)}{\big( (\chi_1-\chi_2)^2 - \frac{n^2}{R^2}\big)\big((\chi_1+\chi_2)^2-\frac{n^2}{R^2}\big)}\\+ \frac{\cot \chi_2 \pi R}{\chi_2}\frac{\big(\chi_2^2-\chi^2_1+3\frac{n^2}{R^2}\big)\big(\chi_2^2+\frac{n^2}{R^2}\big)+
2 \frac{n^2}{R^2}\big(\chi_2^2-\frac{n^2}{R^2}\big)}{\big( (\chi_1-\chi_2)^2 - \frac{n^2}{R^2}\big)\big((\chi_1+\chi_2)^2-\frac{n^2}{R^2}\big)}\bigg]g^{\mu\nu}
\end{multline}

\begin{equation}
\delta \tilde{m}_{n,D2}^{2} = -\frac{1}{2}
\frac{g^2}{2 \pi R}\frac{\zeta(3)}{4 \pi^4 R^2} C_2(G)\ ,\qquad\delta\tilde{m}^{2}_{0,D2} =\frac{1}{4}\frac{g^2}{2 \pi R}\frac{\zeta(3)}{4 \pi^4 R^2} C_2(G)\ .
\end{equation}
Diagram $D3$ yields
\begin{equation}
i\,\Pi^{\mu \nu}_{n,D3} = -\frac{g^2}{4} C_2(G) \delta^{ab}\!\!\int\!\!\! \frac{\deriv^4 p}{(2\pi)^4}
\bigg[\frac{\cot \chi_1 \pi R}{\chi_1}\frac{\big(\chi_1^2-\chi^2_2+\frac{n^2}{R^2}\big)(p_1+p_2)^\mu (p_1+p_2)^\nu }{ \big( (\chi_1-\chi_2)^2 - \frac{n^2}{R^2}\big)\big((\chi_1+\chi_2)^2-\frac{n^2}{R^2}\big)}+ (1\leftrightarrow 2)\bigg]\,,
\end{equation}
\begin{equation}
\delta\tilde{m}_{n,D3}^{2}=0\ , \qquad \delta\tilde{m}_{0,D3}^{2}=-\frac{1}{4}\frac{g^2}{2 \pi R}\frac{\zeta(3)}{4 \pi^4 R^2} C_2(G)\ .
\end{equation}
The ghost loop (diagram $D4$) gives
\begin{multline}
i\,\Pi^{\mu \nu}_{n,D4} = \frac{g^2}{2} C_2(G) \delta^{ab}\int\!\!\! \frac{\deriv^4 p}{(2\pi)^4}p_1^\mu p_2^\nu
\bigg[\frac{\cot \chi_1 \pi R}{\chi_1}\times\\\frac{\big(\chi_1^2-\chi^2_2+\frac{n^2}{R^2}\big)}{\big( (\chi_1-\chi_2)^2 - \frac{n^2}{R^2}\big)\big((\chi_1+\chi_2)^2-\frac{n^2}{R^2}\big)}+ (1\leftrightarrow 2)\bigg]\ ,
\end{multline}
\begin{equation}
\delta \tilde{m}_{n,D4}^{2} =0\ , \qquad \delta \tilde{m}_{0,D4}^{2} =\frac{1}{8}\frac{g^2}{2 \pi R}\frac{\zeta(3)}{4 \pi^4 R^2} C_2(G)\ .
\end{equation}
For the four-vertex (diagram $D5$) one finds
\begin{equation}
i\,\Pi^{\mu \nu}_{n,D5} = -\frac{3}{2} g^2 C_2(G)g^{\mu\nu} \delta^{ab}\int\!\!\! \frac{\deriv^4 p}{(2\pi)^4}
\frac{\cot \chi_1 \pi R}{\chi_1} ,
\end{equation}
\begin{equation}
\delta \tilde{m}_{n,D5}^{2} = \frac{3}{4}\frac{g^2}{2 \pi R}\frac{\zeta(3)}{4 \pi^4 R^2} C_2(G)\ ,
\end{equation}
and for the four-vertex with an $A_5$-loop (diagram $D6$)
\begin{equation}
i\,\Pi^{\mu \nu}_{n,D6} = -\frac{g^2}{2} C_2(G)g^{\mu\nu} \delta^{ab}\int\!\!\! \frac{\deriv^4 p}{(2\pi)^4}
\frac{\cot \chi_1 \pi R}{\chi_1} ,
\end{equation}
\begin{equation}
\delta \tilde{m}_{n,D6}^{2} =\frac{1}{4}\frac{g^2}{2 \pi R}\frac{\zeta(3)}{4 \pi^4 R^2} C_2(G)\ .
\end{equation}
As stated before the sum of all these diagrams is finite.
Also note that the total mass-correction to the zero-mode vanishes, therefore yielding a massless gluon, which is needed
for a realistic low-energy theory.
\newpage

\bibliographystyle{JHEP}
\bibliography{paper}

\providecommand{\href}[2]{#2}\begingroup\raggedright\begin{thebibliography}{10}

\bibitem{Horava:1996qa}
P.~Horava and E.~Witten, {\it Heterotic and type {I} string dynamics from
  eleven dimensions},  {\em Nucl. Phys.} {\bf B460} (1996) 506--524,
  [\href{http://xxx.lanl.gov/abs/hep-th/9510209}{{\tt hep-th/9510209}}].

\bibitem{Antoniadis:1990ew}
I.~Antoniadis, {\it A possible new dimension at a few {T}e{V}},  {\em Phys.
  Lett.} {\bf B246} (1990) 377--384.

\bibitem{Arkani-Hamed:1998nn}
N.~Arkani-Hamed, S.~Dimopoulos, and G.~R. Dvali, {\it Phenomenology,
  astrophysics and cosmology of theories with sub-millimeter dimensions and
  {T}e{V} scale quantum gravity},  {\em Phys. Rev.} {\bf D59} (1999) 086004,
  [\href{http://xxx.lanl.gov/abs/hep-ph/9807344}{{\tt hep-ph/9807344}}].

\bibitem{Witten:1983ux}
E.~Witten, {\it Fermion quantum numbers in {K}aluza-{K}lein theory}, . in
  *Appelquist, T. (Ed.) et al.: modern Kaluza-Klein Theories*, 438-511. (in
  *Shelter Island 1983, proceedings, Quantum Field Theory and the Fundamental
  Problems of Physics*, 227-277) and preprint - Witten, E. (83,Rec.Jan.84) 78
  p.

\bibitem{Dixon:1985jw}
L.~J. Dixon, J.~A. Harvey, C.~Vafa, and E.~Witten, {\it Strings on orbifolds},
  {\em Nucl. Phys.} {\bf B261} (1985) 678--686.

\bibitem{Dixon:1986jc}
L.~J. Dixon, J.~A. Harvey, C.~Vafa, and E.~Witten, {\it Strings on orbifolds.
  2},  {\em Nucl. Phys.} {\bf B274} (1986) 285--314.

\bibitem{Cheng:2002iz}
H.-C. Cheng, K.~T. Matchev, and M.~Schmaltz, {\it Radiative corrections to
  {K}aluza-{K}lein masses},  {\em Phys. Rev.} {\bf D66} (2002) 036005,
  [\href{http://xxx.lanl.gov/abs/hep-ph/0204342}{{\tt hep-ph/0204342}}].

\bibitem{vonGersdorff:2002as}
G.~von Gersdorff, N.~Irges, and M.~Quiros, {\it Bulk and brane radiative
  effects in gauge theories on orbifolds},  {\em Nucl. Phys.} {\bf B635} (2002)
  127--157, [\href{http://xxx.lanl.gov/abs/hep-th/0204223}{{\tt
  hep-th/0204223}}].

\bibitem{Ghilencea:2001ug}
D.~M. Ghilencea and H.-P. Nilles, {\it Quadratic divergences in
  {K}aluza-{K}lein theories},  {\em Phys. Lett.} {\bf B507} (2001) 327--335,
  [\href{http://xxx.lanl.gov/abs/hep-ph/0103151}{{\tt hep-ph/0103151}}].

\bibitem{Delgado:2001ex}
A.~Delgado, G.~von Gersdorff, P.~John, and M.~Quiros, {\it One-loop {H}iggs
  mass finiteness in supersymmetric {K}aluza- {K}lein theories},  {\em Phys.
  Lett.} {\bf B517} (2001) 445--449,
  [\href{http://xxx.lanl.gov/abs/hep-ph/0104112}{{\tt hep-ph/0104112}}].

\bibitem{Contino:2001gz}
R.~Contino and L.~Pilo, {\it A note on regularization methods in
  {K}aluza-{K}lein theories},  {\em Phys. Lett.} {\bf B523} (2001) 347--350,
  [\href{http://xxx.lanl.gov/abs/hep-ph/0104130}{{\tt hep-ph/0104130}}].

\bibitem{Barbieri:2001dm}
R.~Barbieri, L.~J. Hall, and Y.~Nomura, {\it Models of {S}cherk-{S}chwarz
  symmetry breaking in 5{D}: Classification and calculability},  {\em Nucl.
  Phys.} {\bf B624} (2002) 63--80,
  [\href{http://xxx.lanl.gov/abs/hep-th/0107004}{{\tt hep-th/0107004}}].

\bibitem{Arkani-Hamed:1999za}
N.~Arkani-Hamed, Y.~Grossman, and M.~Schmaltz, {\it Split fermions in extra
  dimensions and exponentially small cross-sections at future colliders},  {\em
  Phys. Rev.} {\bf D61} (2000) 115004,
  [\href{http://xxx.lanl.gov/abs/hep-ph/9909411}{{\tt hep-ph/9909411}}].

\bibitem{GrootNibbelink:2001bx}
S.~Groot~Nibbelink, {\it Dimensional regularization of a compact dimension},
  {\em Nucl. Phys.} {\bf B619} (2001) 373--384,
  [\href{http://xxx.lanl.gov/abs/hep-th/0108185}{{\tt hep-th/0108185}}].

\bibitem{Candelas:1984ae}
P.~Candelas and S.~Weinberg, {\it Calculation of gauge couplings and compact
  circumferences from selfconsistent dimensional reduction},  {\em Nucl. Phys.}
  {\bf B237} (1984) 397.

\bibitem{Collins:1984xc}
J.~C. Collins, {\em Renormalization}.
\newblock Cambridge, UK: Univ. Pr., 1984.

\bibitem{Ghilencea:2001bw}
D.~M. Ghilencea, S.~Groot~Nibbelink, and H.~P. Nilles, {\it Gauge corrections
  and {FI}-term in 5{D} {KK} theories},  {\em Nucl. Phys.} {\bf B619} (2001)
  385--395, [\href{http://xxx.lanl.gov/abs/hep-th/0108184}{{\tt
  hep-th/0108184}}].

\bibitem{Gherghetta:2000kr}
T.~Gherghetta and A.~Pomarol, {\it A warped supersymmetric standard model},
  {\em Nucl. Phys.} {\bf B602} (2001) 3--22,
  [\href{http://xxx.lanl.gov/abs/hep-ph/0012378}{{\tt hep-ph/0012378}}].

\bibitem{Arkani-Hamed:2001mi}
N.~Arkani-Hamed, L.~J. Hall, Y.~Nomura, D.~R. Smith, and N.~Weiner, {\it Finite
  radiative electroweak symmetry breaking from the bulk},  {\em Nucl. Phys.}
  {\bf B605} (2001) 81--115,
  [\href{http://xxx.lanl.gov/abs/hep-ph/0102090}{{\tt hep-ph/0102090}}].

\bibitem{Barbieri:2002uk}
R.~Barbieri, G.~Marandella, and M.~Papucci, {\it Breaking the electroweak
  symmetry and supersymmetry by a compact extra dimension},  {\em Phys. Rev.}
  {\bf D66} (2002) 095003, [\href{http://xxx.lanl.gov/abs/hep-ph/0205280}{{\tt
  hep-ph/0205280}}].

\bibitem{Das:1997gg}
A.~K. Das, {\em Finite temperature field theory}.
\newblock Singapore, Singapore: World Scientific, 1997.

\bibitem{Hosotani:1983xw}
Y.~Hosotani, {\it Dynamical mass generation by compact extra dimensions},  {\em
  Phys. Lett.} {\bf B126} (1983) 309.

\bibitem{Georgi:2000ks}
H.~Georgi, A.~K. Grant, and G.~Hailu, {\it Brane couplings from bulk loops},
  {\em Phys. Lett.} {\bf B506} (2001) 207--214,
  [\href{http://xxx.lanl.gov/abs/hep-ph/0012379}{{\tt hep-ph/0012379}}].

\bibitem{Delgado:2002xf}
A.~Delgado, G.~von Gersdorff, and M.~Quiros, {\it Brane-assisted
  {S}cherk-{S}chwarz supersymmetry breaking in orbifolds},  {\em JHEP} {\bf 12}
  (2002) 002, [\href{http://xxx.lanl.gov/abs/hep-th/0210181}{{\tt
  hep-th/0210181}}].

\bibitem{Peskin:1995ev}
M.~E. Peskin and D.~V. Schroeder, {\it An introduction to quantum field
  theory}, . Reading, USA: Addison-Wesley (1995) 842 p.

\bibitem{Magnus:1948}
W.~Magnus and F.~Oberhettinger, {\em Formeln und S{\"a}tze f{\"u}r die
  speziellen Funktionen der mathematischen Physik}.
\newblock Berlin, Germany: Springer, 1948.

\end{thebibliography}\endgroup
\end{document}